\documentclass[10pt, journal, letterpaper]{IEEEtran}
\usepackage[utf8]{inputenc}
\usepackage{amsmath}
\usepackage{amsthm}
\usepackage{amssymb}
\usepackage{mathtools}
\usepackage{enumerate}
\usepackage{bbm}
\usepackage{algorithm}
\usepackage{algorithmic}
\usepackage{epsfig}
\usepackage{xcolor}
\usepackage{graphicx}
\usepackage{caption}
\usepackage{subcaption}
\usepackage{cases}
\usepackage{url}
\usepackage{cite}
\usepackage{mleftright}
\usepackage{xparse}
\usepackage{multirow}
\usepackage{xr}
\usepackage{comment}
\usepackage{float} 
\usepackage{multicol}

\DeclareMathOperator*{\argmin}{arg\,min}

\usepackage{soul}

\newtheorem{lemma}{\bf Lemma}[section]

\begin{document}

\title{Timely Trajectory Reconstruction in Finite \\ Buffer Remote Tracking Systems}

% \author{Sunjung Kang,~\IEEEmembership{Member,~IEEE,}, Christopher G. Brinton,~\IEEEmembership{Member,~IEEE,}, Vishrant Tripathi
\author{Sunjung~Kang,
        Vishrant Tripathi,
        and Christopher G. Brinton% <-this % stops a space
    \thanks{S. Kang, V. Tripathi and C. G. Brinton are with Elmore Family School of Electrical and Computer Engineering, Purdue University, IN 47907, USA.
    E-mail: \{kang392,tripathv,cgb\}@purdue.edu
    }
    \thanks{An earlier version of this paper was presented at the International Symposium on Modeling and Optimization in Mobile, Ad Hoc, and Wireless Networks (WiOpt), 2025~\cite{kang2025technical}.}
}


\maketitle
\begin{abstract}
    Remote tracking systems play a critical role in applications such as IoT, monitoring, surveillance and healthcare. In such systems, maintaining both real-time state awareness (for online decision making) and accurate reconstruction of historical trajectories (for offline post-processing) are essential. While the Age of Information (AoI) metric has been extensively studied as a measure of freshness, it does not capture the accuracy with which past trajectories can be reconstructed. In this work, we investigate reconstruction error as a complementary metric to AoI, addressing the trade-off between timely updates and historical accuracy. Specifically, we we consider a remote tracking system where the source evolves as a Wiener process and examine three policies, each prioritizing different aspects of information management: Keep-Old, Keep-Fresh, and our proposed Inter-arrival-Aware dropping policy. We compare these policies in terms of impact on both AoI and reconstruction error in a remote tracking system with a finite buffer. Through theoretical analysis and numerical simulations of queueing behavior, we demonstrate that while the Keep-Fresh policy minimizes AoI, it does not necessarily minimize reconstruction error. In contrast, our proposed Inter-arrival-Aware dropping policy dynamically adjusts packet retention decisions based on generation times, achieving a balance between AoI and reconstruction error. Our results provide key insights into the design of efficient buffer management policies for resource-constrained IoT networks.
\end{abstract}

\section{Introduction}
% \RED{footnote - conference version}

With the emergence of the Internet of Things (IoT), remote tracking systems have been gaining much attention in monitoring for environmental, agricultural, urban and personal healthcare applications~\cite{atzori2010internet}. These systems rely on sensors transmitting time-varying data packets to remote monitors for real-time tracking and analysis of objects and their trajectories. Maintaining the freshness and accuracy of updates is crucial for ensuring effective system performance in such applications.

The Age of Information (AoI) has been widely studied as a measure of information freshness, quantifying the time elapsed since the last update was generated~\cite{sun2017update, pan2022age}. The AoI has become a cornerstone for analyzing timeliness in applications where real-time updates are critical. However, many applications also require accurate trajectory reconstruction alongside real-time monitoring. For instance, in a surveillance application, real-time monitoring enables immediate detection, while trajectory reconstruction helps analyze long-term movement patterns. Similarly, in healthcare, wearable devices provide real-time alerts, while reconstructing activity trajectories for long-term fitness tracking and personalized recommendations. To capture these dual objectives, we consider employing reconstruction error as a complementary metric to AoI.

Balancing freshness and reconstruction accuracy, however, is particularly challenging in resource-constrained IoT environments. One major challenge is the limited buffer size at the transmitter, which necessitates making decisions on which packets to retain and which to discard. Prior studies have shown that age-optimal policies retain only the freshest updates, requiring minimal buffering~\cite{costa2016age}. However, this approach inevitably leads to the discarding of older packets that are crucial for reconstructing trajectories accurately.

In this paper, we investigate how packet-dropping policies and finite buffer sizes impact both the AoI and the reconstruction accuracy in remote tracking systems. We compare three policies: (i) Keep-Old, which prioritizes historical packets; (ii) Keep-Fresh, which minimizes AoI by retaining the most recent updates; and (iii) our Inter-arrival-Aware policy, which dynamically balances both objectives based on system conditions.

\subsection{Related Works}

A key challenge in AoI research is optimizing sampling and scheduling to minimize information staleness~\cite{sun2017update,yao2022age,ozel2021intermittent,pan2022age,liu2024learning,costa2016age,bedewy2017age,talak2019age,kosta2019queue,kadota2018scheduling,talak2018scheduling,bedewy2021optimal,tripathi2023wiswarm,tripathi2024whittle,ramakanth2024monitoring,bedewy2021low,tripathi2023fresh,rao2023age}. This involves determining the ideal moments to sample data and transmit updates in a way that keeps the information at the receiver as timely as possible while adhering to system constraints such as energy, bandwidth, and communication delays. Many works in this domain assume the ability to generate updates at will, and explore trade-offs under random transmission times~\cite{sun2017update}, unreliable channels with two-way delays~\cite{yao2022age}, intermittent transmissions~\cite{ozel2021intermittent}, and energy constraints~\cite{pan2022age}, with learning-based solutions developed for dynamic environments~\cite{liu2024learning}.

In contrast, in settings where updates arrive randomly rather than at will, researchers have studied queue management and service disciplines to optimize AoI. The impact of queue discipline such as FCFS or LCFS on AoI has been analyzed, which shows that different disciplines can significantly affect information freshness~\cite{costa2016age}. This line of research has been extended by analyzing the impact of preemption policies~\cite{bedewy2017age} and packet replacement strategies~\cite{kosta2019queue}, both shown to be effective in further improving AoI performance. The trade-off between minimizing AoI and controlling delay has also been highlighted, which shows that aggressive freshness optimization can lead to increased queuing delays~\cite{talak2019age}. While these works offer a solid foundation for understanding AoI dynamics in systems with limited control over update generation, they typically prioritize freshness without considering downstream estimation performance or the value of historical information.

In wireless networks, especially in multi-source scenarios, scheduling plays a vital role in minimizing AoI. Centralized scheduling approaches have introduced policies such as Greedy, Max-Weight, Whittle’s Index~\cite{kadota2018scheduling}, and Drift-Plus-Penalty~\cite{kadota2019scheduling} to optimize AoI while satisfying throughput constraints. Extensions include age-based and virtual-queue-based scheduling for unknown and time-varying channels~\cite{talak2018scheduling}, Maximum Age First (MAF) strategies that separate sampling and scheduling~\cite{bedewy2019age}, and joint sampling-scheduling designs~\cite{bedewy2021optimal}. Whittle index-based scheduling has been generalized to broader AoI cost functions~\cite{tripathi2024whittle}, and packet selection from buffers has been explored for inference-aware settings~\cite{shisher2024timely}. In addition, policies for multi-hop~\cite{bedewy2016optimizing} and multi-server systems, as well as randomized scheduling under stochastic arrivals~\cite{kadota2019minimizing}, have demonstrated improvements in AoI. AoI minimization under monitoring error objectives for correlated sources modeled as Wiener processes has also been studied~\cite{ramakanth2024monitoring}.

Decentralized scheduling has also received attention due to its practicality in large-scale and uncoordinated systems. Proposed approaches include asynchronous sleep-wake scheduling~\cite{bedewy2021low}, reinforcement learning-based schemes for vehicular networks~\cite{qin2021distributed}, and Fresh-CSMA protocols that adapt backoff times based on local AoI~\cite{tripathi2023fresh}. Estimation-aware metrics such as Trackability-aware AoI (TAoI)\cite{choudhury2021joint} and new metrics like Age of Broadcast (AoB) and Age of Collection (AoC)\cite{rao2023age} have been proposed. In random access networks, decentralized policies using age-gain~\cite{chen2020age}, optimized backoff for CSMA~\cite{maatouk2019minimizing}, and stochastic hybrid system models~\cite{wang2024analytical} have been introduced to address collisions, finite buffers, and trade-offs with throughput. While these studies offer valuable insights into network-wide AoI optimization, they primarily focus on freshness across multiple sources, which differs from our setting centered on single-source reconstruction under buffer constraints.

Research in remote estimation has addressed a wide range of system models and constraints to improve state estimation accuracy. In single-source settings, optimal event-triggered sampling policies have been developed for Wiener~\cite{nar2014sampling,sun2017remote} and Ornstein-Uhlenbeck processes~\cite{ornee2019sampling}, with extensions incorporating random delays and channel dynamics. AoI has been introduced into estimation frameworks to analyze freshness-performance trade-offs in settings like random access channels~\cite{chen2021real} and industrial wireless sensor networks~\cite{chen2020mse}. Further studies have proposed scheduling and retransmission policies to reduce estimation error in systems with packet drops, LTI dynamics, or fading channels~\cite{huang2020real,yun2018optimal,gao2015optimal1,gao2015optimal2}. In multi-source scenarios, centralized and decentralized strategies address interference and resource sharing~\cite{han2017optimal,wu2018optimal,zhang2021distributed}, using tools like correlated equilibria~\cite{ding2017multi,chakravorty2019remote}, Whittle index~\cite{ornee2023context,ornee2023whittle}, and coded or quantized updates~\cite{arafa2020timely,banawan2022timely}. One-bit update strategies~\cite{kang2023remote} and comparisons of centralized versus decentralized architectures~\cite{kang2023comparison} have also been explored to reduce communication cost while maintaining estimation quality.

While most prior work focuses on tracking the current source state, a growing line of research recognizes that AoI alone is insufficient for inference, as it neglects the source’s statistical characteristics. Recent efforts have developed alternative metrics and update strategies that align more closely with estimation objectives. Examples include communication-learning co-design frameworks for feature selection and scheduling~\cite{shisher2023learning}, context-aware status updating policies~\cite{ornee2023context}, and the uncertainty-of-information scheduling paradigm~\cite{chen2022uncertainty,chen2023index,chen2024optimal}, which quantifies the informativeness of packets beyond their recency. These approaches collectively highlight the need to assess how well received information supports the intended inference goals of the system.

Our work is aligned with this broader perspective, but introduces a distinct angle by focusing on reconstructing past state trajectories. We consider remote tracking systems with finite buffer constraints, where the goal is not only to maintain update freshness but also to retain packets that are informative for rebuilding historical sequences. By analyzing packet dropping strategies, we capture the trade-off between delivering timely updates and preserving valuable past information, which highlights limitations of AoI-centric approaches in applications requiring retros pective insight.

\subsection{Contributions}

In this paper, we consider a remote tracking system where packets are randomly generated, queued, and transmitted from a source to a receiver. Unlike conventional AoI-based policies that focus solely on maintaining information freshness, we aim to balance two objectives: (i) minimizing AoI for real-time monitoring, and (ii) reducing reconstruction error for accurate trajectory reconstruction. Unlike remote estimation, which minimizes immediate state errors, our approach captures cumulative historical accuracy in order to improve the reconstruction of past states.

Our contributions can be summarized as follows: 
\begin{itemize}
    \item We formulate the problem of balancing real-time monitoring and trajectory reconstruction in remote tracking systems. Using reconstruction error alongside AoI, we capture the trade-off between ensuring timely updates for responsiveness and retaining older data for accurate historical reconstruction.
    %We formulate the problem of balancing real-time monitoring and accurate trajectory reconstruction in remote tracking systems. Using the {\it reconstruction error} metric alongside the AoI, we address the challenge of prioritizing updates under resource constraints, such as limited buffer size and transmission capacity. This problem captures the trade-offs between ensuring timely updates for immediate responsiveness and retaining older data necessary for reconstructing trajectories, which are critical for applications like surveillance and healthcare.
    \item We analyze the average peak age and reconstruction error under two fixed packet-dropping policies: (i) Keep-Old, which prioritizes historical data, and (ii) Keep-Fresh, which retains the most recent packets. Our results show that Keep-Old generally performs worse in both metrics, with Keep-Fresh providing better overall performance.
    %We analyze the average peak age and average reconstruction error of two fixed packet-dropping policies: Keep-Old Policy, which prioritizes retaining historical packets, and Keep-Fresh Policy, which prioritizes fresh packets. Our analysis shows that the Keep-Old policy performs poorly compared to the Keep-Fresh policy in both age and reconstruction error under most scenarios.   
    \item To further minimize reconstruction accuracy while maintaining freshness, we introduce (iii) the Inter-arrival-Aware (IaA) dropping policy, which dynamically selects packets based on their generation times. By considering temporal diversity, this policy improves reconstruction accuracy without significantly compromising on age performance. 
    \item We then introduce (iv) the Threshold-based Inter-arrival-Aware dropping policy, which extends the IaA approach by incorporating a threshold mechanism. This modification enables a more balanced trade-off between reconstruction accuracy and freshness. 
    \item Finally, through numerical experiments, we evaluate how different dropping policies and buffer sizes influence AoI and reconstruction error. Our results provide insights into designing efficient update management strategies under resource constraints. 
\end{itemize}

\section{System Model and Problem Formulation} \label{sec:system_model}

\begin{figure}[t]
    \centering
    \includegraphics[width=\linewidth]{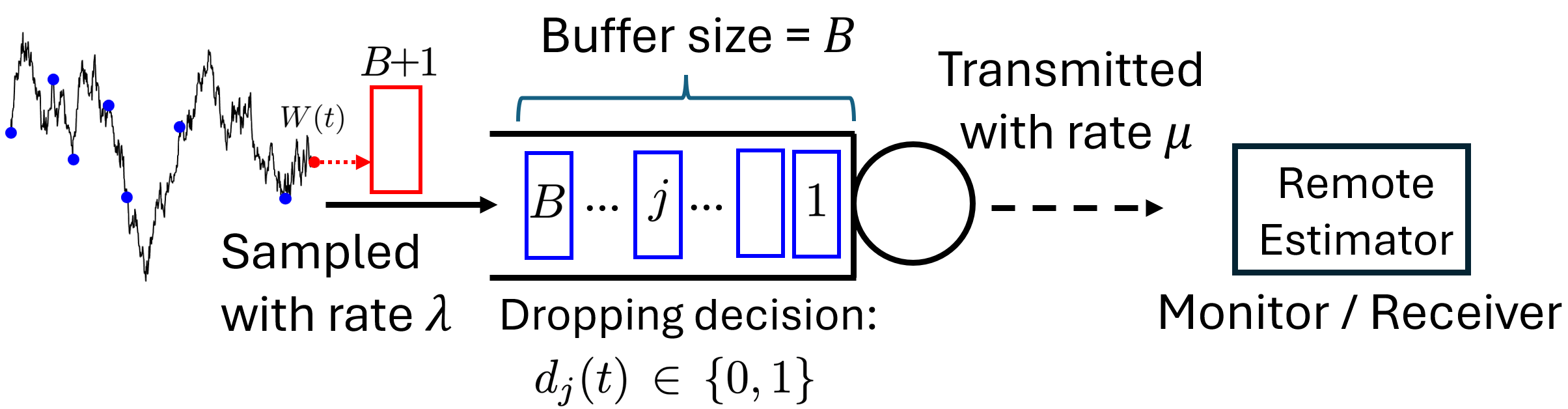}
    \caption{System model.}
    \label{fig:system_model}
\end{figure}

We consider a network where a sensor generates packets containing the state of a dynamic source and temporarily stores these packets in a finite buffer before transmission to an estimator, as illustrated in Fig.~\ref{fig:system_model}. Packets in the queue are served under a non-preemptive Last-Come-First-Serve (LCFS) queuing discipline. The estimator uses the received packets for both real-time tracking of the current state and offline reconstruction of past trajectories. Due to buffer constraints and limited transmission resources, the transmitter must make packet-dropping decisions that consider both the freshness (real-time value) and historical importance of each packet.

\subsection{System Model}

The state of the source, denoted as $W(t)$, evolves according to a Wiener process, a continuous-time stochastic model characterized by independent increments. Specifically, increments over non-overlapping intervals are independent and normally distributed with mean zero and variance proportional to the length of the interval. Update packets containing state measurements are generated according to a Poisson process with rate $\lambda > 0$. Let $t_j$ denote the generation time of packet $j$. Each update packet $j$ contains the state measurement $W(t_j)$ at its generation time and is stored in a finite buffer of size $B$. Packets in the LCFS queue are served non-preemptively, i.e., ongoing transmissions are never interrupted.

When the buffer is full and a new packet arrives, the transmitter must decide whether to drop the new packet or replace an existing packet already in the buffer. Let $d_j(t) \in {0,1}$ denote the dropping decision at time $t$, where $j \in {1,...,B,B+1}$. The indices $j=1$ and $j=B$ correspond to packets currently at the head and tail of the buffer, respectively, while $j=B+1$ represents the newly arrived packet. If $d_{B+1}(t)=1$, the newly arrived packet is dropped. Conversely, if $d_j(t)=1$ for $j\in \{1,...,B\}$, the corresponding packet in the buffer is dropped, and the new packet is placed at the tail of the buffer. This dropping decision must balance the freshness of real-time tracking data and the packet’s historical importance for reconstructing past trajectories.

% At the receiver, the received packets are used to reconstruct the source’s historical trajectory. However, packet drops and a limited sampling rate create information gaps, introducing uncertainty in offline reconstruction over a long horizon. Thus, the receiver must estimate the state of the source during periods for which no direct observations are available. 

\subsection{Age of Information}

\begin{figure}
    \centering
    \includegraphics[width=0.9\linewidth]{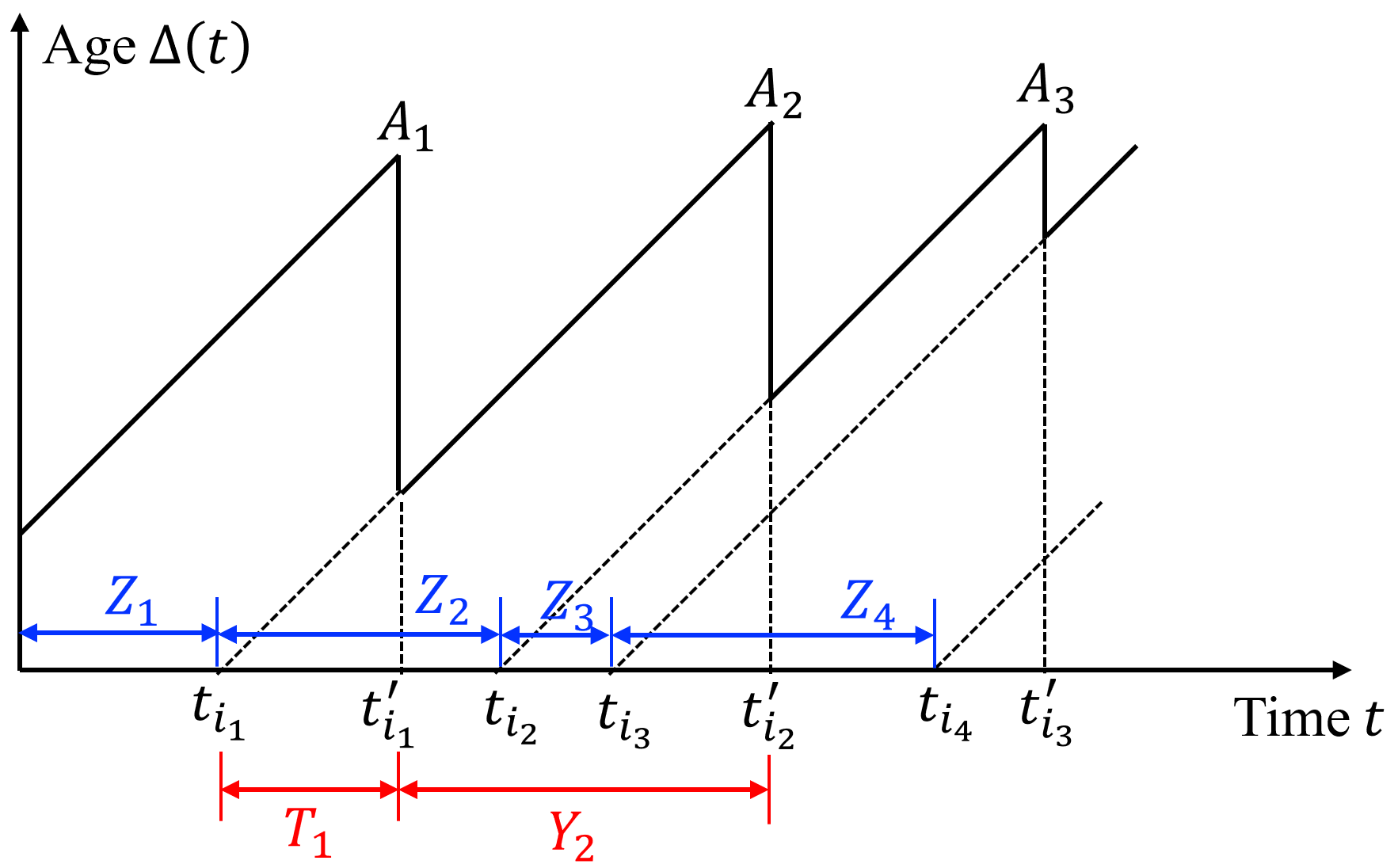}
    \caption{Sample path of age $\Delta(t)$, where packet $2$ is delivered after packet $3$, making it stale and thus not reducing the age. Consequently $i_2 = 3$.}
    \label{fig:age_evolution}
\end{figure}

Packets serve the dual roles of real-time state tracking and historical reconstruction.  The {\it Age of Information (AoI)} $\Delta_j(t)$ for update packet $j$ is a key measure of how fresh a packet is, where $\Delta_j(t) = t - t_j$.  Lower AoI indicates fresher information, which makes it more relevant for real-time tracking. The AoI $\Delta(t)$ for the system at time $t$ is defined as the age of the freshest delivered packet: 
\begin{equation} 
    \Delta(t) = t - \max_{\{i : t'_{i} \le t\}} t_{i}, \end{equation} 
where $t'_{i}$ is the delivery time of update packet $i$. We call that packet $j$ is fresh at time $t$ if $\Delta_j(t) < \Delta(t)$.

We note that not all generated packets are delivered in the order of their generation time. Let $i_k$ denote the index of the $k^{\text{th}}$ fresh packet. Further, we let $A_k$ denote the $k^{\text{th}}$ peak age, which is the maximum value of the age immediately before the $k^{th}$ fresh packet is received at the receiver: 
\begin{equation} 
    A_k = \Delta((t'_{i_k})^-), 
\end{equation} 
where $f(t^-) := \lim_{s\rightarrow t^-}f(s)$.

Fig.~\ref{fig:age_evolution} shows a sample path of age $\Delta(t)$ evolution over time. Let $T_{i_k}:=t'_{i_k}-t_{i_k}$ denote the system time of packet $i_k$, representing the total time from the generation to the delivery of the $k^{\text{th}}$ fresh packet. Let $Y_k:=t'_{i_k}-t'_{i_{k-1}}$ denote the inter-delivery time, which is the time interval between the deliveries of packets $i_{k-1}$ and $i_k$. Then, the $k^{th}$ peak age $A_k$ can be expressed as
\begin{equation}
    A_{k} = T_{i_{k-1}} + Y_k. 
\end{equation}
The time-average peak age is then given by
\begin{equation}
    \Bar{A} = \lim_{K\rightarrow\infty} \frac{1}{K} \sum_{k=1}^K A_k = \lim_{K\rightarrow\infty} \frac{1}{K} \sum_{k=1}^K (T_{i_{k-1}} + Y_k).
\end{equation}
Due to the ergodicity of the process of packet generations and deliveries, the time-average peak age can be expresssed as
\begin{equation}
    \Bar{A} = \mathbb{E}[T_{i_{k-1}}] + \mathbb{E}[Y_k]. \label{eq:peak_age_expression}
\end{equation}

\subsection{Reconstruction Error} \label{subsec:RE}

In addition to real-time tracking, the receiver aims to reconstruct the historical trajectory of the source’s state using the received packets. However, due to packet drops and the limited sampling rate, there is often missing information between the times at which packets are generated. These gaps introduce uncertainty into the estimation of the source's state over time.

To quantify the accuracy of this reconstruction, we define the reconstruction error (RE) as the mean squared deviation between the true source trajectory and its estimate. Let $\hat{W}(t)$ denote the reconstruction of the true state $W(t)$. Let $\tilde{T}$ denote the latest delivery time before $T$, i.e.,
\begin{equation}
    \tilde{T} = \max\{t_i' ~:~ t_i' \le T\},
\end{equation}
where $t_i'$ is the delivery time of packet $i$. We use $\tilde{T}$ to indicate the most recent point up to which the reconstruction is evaluated based on the available delivered data. The RE over a time interval $[0,T]$ is given by:
\begin{equation}
    \text{RE}(T) = \frac{1}{\tilde{T}}\int_{0}^{\tilde{T}} (W(s)-\hat{W}(s))^2 ~ds.
\end{equation}

Since our goal is to reconstruct the trajectory offline using observations at $t_i$ and $t_{i+1}$, we adopt the Linear Minimum Mean Square Error (LMMSE) estimator. For Wiener processes, which are Gaussian with continuous paths and independent increments, the LMMSE estimate of $W(t)$ for $t \in (t_i, t_{i+1})$ is the conditional expectation given the two observations $\hat{W}(t) = \mathbb{E}[W(t) | W(t_i), W(t_{i+1})]$. This conditional expectation reduces to linear interpolation since the covariance of the Wiener process grows linearly with time~\cite{durrett2019probability}:
\begin{equation} 
\hat{W}(t) = W(t_i) + \frac{t - t_i}{t_{i+1} - t_i} \left(W(t_{i+1}) - W(t_i)\right)
\end{equation}
for $t_{i}<t<t_{i+1}$.

Under this reconstruction scheme, the RE can be analytically expressed. Let $n(T)$ denote the number of packets delivered until time $T$. Then, the RE for the Wiener process becomes~\cite{kang2024balancing}:
\begin{equation}
    \text{RE}(T) = \frac{1}{\tilde{T}} \sum_{k=1}^{n(\tilde{T})} \frac{(t_{i+1}-t_{i})^2}{6}. \label{eq:RE}
\end{equation}
For completeness of the derivation of the LMMSE estimator and the analytical expression for the reconstruction error, we refer to \ref{appendix:Wiener}. We can rewrite \eqref{eq:RE} as:
\begin{equation}
    \text{RE}(T) = \frac{n(\tilde{T})}{\tilde{T}} \cdot \frac{1}{n(\tilde{T})} \sum_{i=1}^{n(\tilde{T})} \frac{(t_{i+1}-t_{i})^2}{6}.
\end{equation}
As $T \rightarrow \infty$ (and consequently $\tilde{T} \rightarrow \infty$), the delivery rate converges to the effective arrival rate $\lambda_{\text{eff}} = \lambda (1 - P_L)$, where $P_L$ is the packet loss probability. Let $Z_i = t_{i} - t_{i-1}$ denote the inter-generation time. Due to ergodicity, we obtain:
\begin{equation}
    \lim_{T\rightarrow\infty} \frac{1}{n(\tilde{T})} \sum_{i=1}^{n(\tilde{T})} (t_{i+1}-t_{i})^2 = \mathbb{E}[Z_i^2].
\end{equation}
Hence, the long-term average reconstruction error is:
\begin{equation}
    \overline{\text{RE}} := \lim_{T\rightarrow\infty}  \text{RE}(T) = \frac{\lambda_{\text{eff}}\mathbb{E}[Z_i^2]}{6}.
\end{equation}

To evaluate $\lambda_{\text{eff}}$, we consider the queue dynamics. Let $\pi_i$ be the steady-state probability that the system contains $i$ packets, for $i \in \{0,1,...,B+1\}$, in an $M/M/1/B+1$ system. Then, we have
\begin{equation}
    \pi_i = \frac{\rho^i}{\sum_{j=0}^{B+1} \rho^j}, \quad \text{where } \rho = \frac{\lambda}{\mu}, \label{eq:steady_prob}
\end{equation}
where $\rho = \frac{\lambda}{\mu}$ is the traffic intensity. Note that, in an $M/M/1/B+1$ queue, the packet loss probability $P_L$ is equivalent to the probability that the system is busy, i.e., $P_L = \pi_{B+1}$. Thus, the effective arrival rate is given by
\begin{equation}
    \lambda_{\text{eff}} = \lambda(1 - \pi_{B+1}) = \frac{\lambda (1 - \rho^{B+1})}{1 - \rho^{B+2}}. \label{eq:effective_arrival_rate}
\end{equation}

It is important to note that while the expected inter-arrival time $\mathbb{E}[Z_i] = 1 / \lambda_{\text{eff}}$ is independent of the dropping policy, the second moment $\mathbb{E}[Z_i^2]$ is policy-dependent. This is because dropping decisions influence which packets are ultimately delivered and thus alter the distribution of inter-arrival intervals between successfully delivered packets. Consequently, different packet-dropping strategies affect the reconstruction performance, as captured by the RE. This highlights the importance of designing policies that consider both timeliness and the informativeness of packet timings for effective trajectory reconstruction.

% Let $\pi_i$ be the steady state probability of queue length being $i$ for $i\in\{0,1,...,B+1\}$, which is given by 
% \begin{equation}
%     \pi_i = \frac{\rho^i}{\sum_{j=0}^{B+1} \rho^j} \text{ for } i = 0,1,\cdots,B+1, \label{eq:steady_prob}
% \end{equation}
% where $\rho = \frac{\lambda}{\mu}$ is the traffic intensity. Note that, in an $M/M/1/B+1$ queue, the packet loss probability $P_L$ is equivalent to the probability that the system is busy, i.e., $P_L = \pi_{B+1}$. Thus, the effective arrival rate is given by
% \begin{equation}
%     \lambda_{\text{eff}} = \lambda(1-\pi_{B+1}) = \frac{\lambda (1-\rho^{B+1})}{1-\rho^{B+2}}. \label{eq:effective_arrival_rate}
% \end{equation}

% The expected inter-arrival time $\mathbb{E}[Z_i]$ is independent of the dropping policy, which is given by $\frac{1}{\lambda_{\text{eff}}}$. However, the second moment $\mathbb{E}[Z^2_i]$ depends on the specific packet-dropping policy. Different policies affect the queueing dynamics, particularly the distribution of inter-arrival times between successfully delivered packets. These inter-arrival times are influenced by how packets are prioritized or dropped in the system. As a result, the second moment $\mathbb{E}[Z_i^2]$ varies under different policies, which leads to distinct impacts on system performance metrics such as the reconstruction error. This highlights the critical role of packet-dropping policies in determining the overall system behavior and performance.

\section{$M/M/1/2$ Queuing Systems} \label{sec:MM12}

In this section, we analyze an $M/M/1/2$ queue, where both inter-arrival times and services times follow a memoryless, exponential distribution. The system consists of a single server and one waiting slot, focusing on the Keep-Old and Keep-Fresh policies. While prior studies have focused on the AoI as a performance measure, we extend this analysis to include reconstruction error. In a non-preemptive $M/M/1/2$ system, all delivered packets are fresh ($i_k = k$), as the buffer holds at most one packet, ensuring that the transmitted packet is always the freshest available.

\subsection{Keep-Old Policy} \label{subsec:single_keep_old}

\begin{figure}
    \centering
    \includegraphics[width=\linewidth]{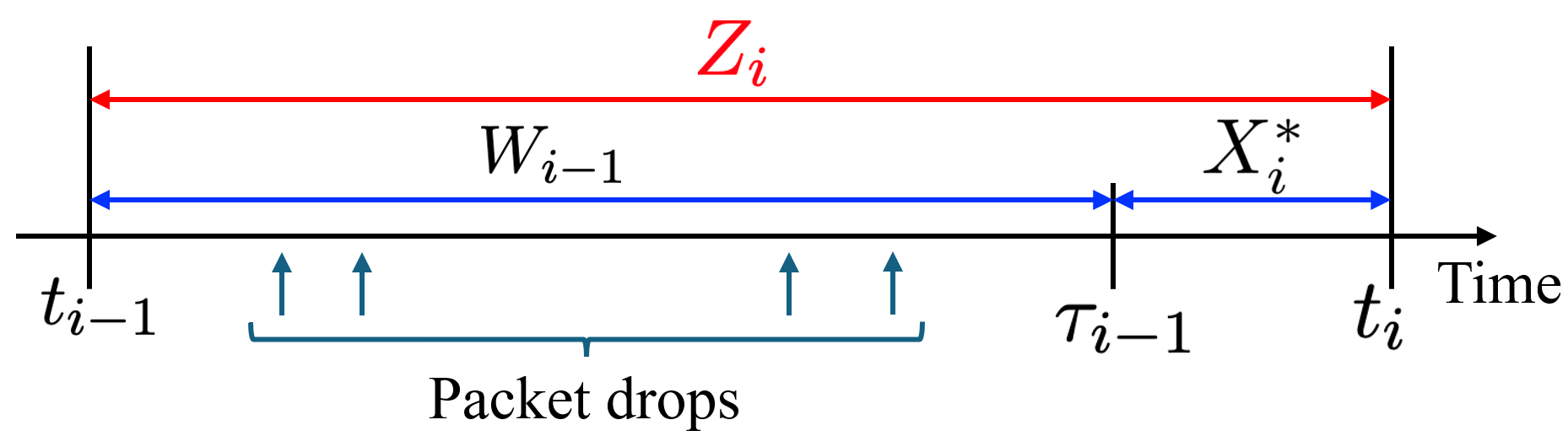}
    \caption{Sample path of packet arrivals and departures under the Keep-Old policy.}
    \label{fig:MM12_keep_old_time_line}
\end{figure}

The Keep-Old policy always retains older packets in the buffer and drops new arrivals when the buffer is full, i.e., $d_1(t) = 0$ and $d_2(t) = 1$. This strategy preserves historical information but often leads to outdated packets being transmitted, which increases the AoI. The average peak age $\bar{A}$ under the Keep-Old policy is analyzed in~\cite{costa2016age}:
\begin{equation}
    \bar{A}_{\text{Keep-Old}} = \frac{1}{\lambda}+\frac{3}{\mu}-\frac{2}{\lambda+\mu} \rightarrow \frac{3}{\mu} ~\text{ as } \lambda\rightarrow\infty.
\end{equation}
% which asymptotically approaches:
% \begin{equation}
%     \bar{A}_{\text{Keep-Old}}\rightarrow \frac{3}{\mu} ~\text{ as } \lambda\rightarrow\infty.
% \end{equation}

% We now derive the average reconstruction error under this policy. Since only one packet can wait in the system, the inter-arrival time between two delivered packets depends on both the time a packet waits and the arrival time of the next accepted packet. Fig.~\ref{fig:MM12_keep_old_time_line} illustrates this dynamic.

% \begin{lemma} \label{lemma:Keep_old_rec_error} \it
%     The long-term average reconstruction error under the Keep-Old policy is:
%     \begin{equation}
%          \overline{RE}_{\text{Keep-Old}} =\frac{(\lambda+\mu)(\lambda^2+\mu^2)}{3\lambda\mu(\lambda^2+\lambda\mu+\mu^2)} \rightarrow \frac{1}{3\mu} \text{ as } \lambda \to \infty.
%     \end{equation}
% \end{lemma}

% The full derivation is given in Appendix~\ref{appendix:Keep_old_rec_error}, but the key insight is that RE is driven by the second moment of inter-delivery times, which can be computed using the memoryless property of the exponential distributions and the PASTA principle.

We now derive the average reconstruction error under this policy. As discussed in Section~\ref{subsec:RE}, the reconstruction error depends on the second moment $\mathbf{E}[Z_i^2]$ of the inter-arrival times between successfully delivered packets. The inter-arrival time $Z_i$ consists of two independent components: the waiting time $W_{i-1}$, which depends on whether the packet finds the system busy upon arrival, and the service-to-arrival time $X^*_i$, the interval between the service start of packet $i-1$ and the arrival of packet $i$ (see Fig.~\ref{fig:MM12_keep_old_time_line}). This leads to the following expression:
\begin{equation}
     \mathbb{E}[Z_i^2] = \mathbb{E}[W_{i-1}^2] + 2\mathbb{E}[W_{i-1}]\mathbb{E}[X^*_i] + \mathbb{E}[(X^*_i)^2].
\end{equation}

Using memoryless properties of inter-arrival times and service times and the PASTA property, we can obtain
\begin{equation}
    \begin{split}
         \mathbb{E}[Z_i^2] = \frac{2}{\mu^2} + \frac{2}{\lambda^2}.
    \end{split}   
\end{equation}
Combining with (\ref{eq:effective_arrival_rate}), we present the following lemma.
%a lemma that provides the long-term average reconstruction error $\overline{RE}_{\text{Keep-Old}}$ under the Keep-Old policy.
\begin{lemma} \label{lemma:Keep_old_rec_error} \it
    The long-term average reconstruction error under the Keep-Old policy in an M/M/1/2 queuing system is given by
    \begin{equation}
         \overline{RE}_{\text{Keep-Old}} =\frac{(\lambda+\mu)(\lambda^2+\mu^2)}{3\lambda\mu(\lambda^2+\lambda\mu+\mu^2)}.
    \end{equation}
\end{lemma}
\vspace{0.1in}

From this lemma, we can see that the average reconstruction error asymptotically approaches $\frac{1}{3\mu}$.
% \begin{equation}
%     \textstyle \overline{RE}_{\text{Keep-Old}}\rightarrow \frac{1}{3\mu} ~\text{ as }\lambda\rightarrow\infty.
% \end{equation}
We provide a detailed proof in Appendix~\ref{appendix:Keep_old_rec_error}.

\subsection{Keep-Fresh Policy} \label{subsec:MM12_keep_fresh}
The Keep-Fresh policy replaces the existing packet in the buffer when a new arrival occurs, i.e, $d_1(t) = 1$ and $d_2(t) = 0$, ensuring that only the freshest updates are retained. This policy minimizes AoI, making it ideal for real-time applications. The average peak AoI under this policy is given by~\cite{costa2016age}
\begin{equation}
    \bar{A}_{\text{Keep-Fresh}} = \frac{1}{\mu} + \frac{\lambda}{(\lambda+\mu)^2}+\frac{1}{\lambda}+\frac{1}{\mu}\frac{\lambda}{\lambda+\mu}, 
\end{equation}
which asymptotically approaches:
\begin{equation}
    \bar{A}_{\text{Keep-Fresh}} \rightarrow \frac{2}{\mu} ~\text{ as } \lambda\rightarrow\infty.
\end{equation}

\begin{figure}
    \centering
    \begin{subfigure}[t]{\linewidth}
        \centering
        \includegraphics[width=\linewidth]{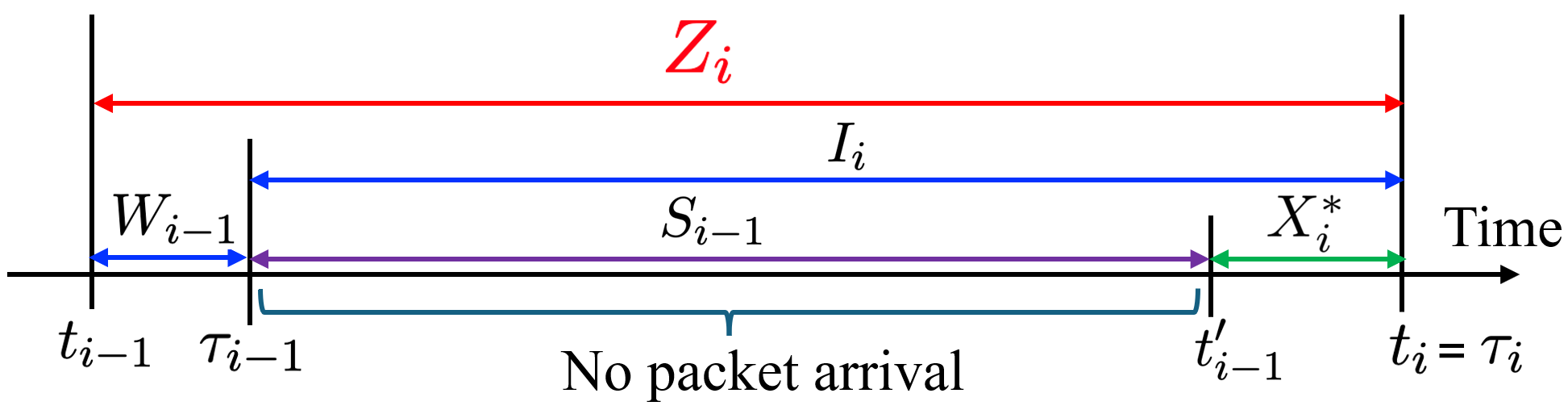}
        \caption{No arrival occurs during the service time of packet $i_{k-1}$.}
        \label{subfig:replacment_time_line1}
    \end{subfigure}
    \hfill
    \begin{subfigure}[t]{\linewidth}
        \centering     
        \includegraphics[width=\linewidth]{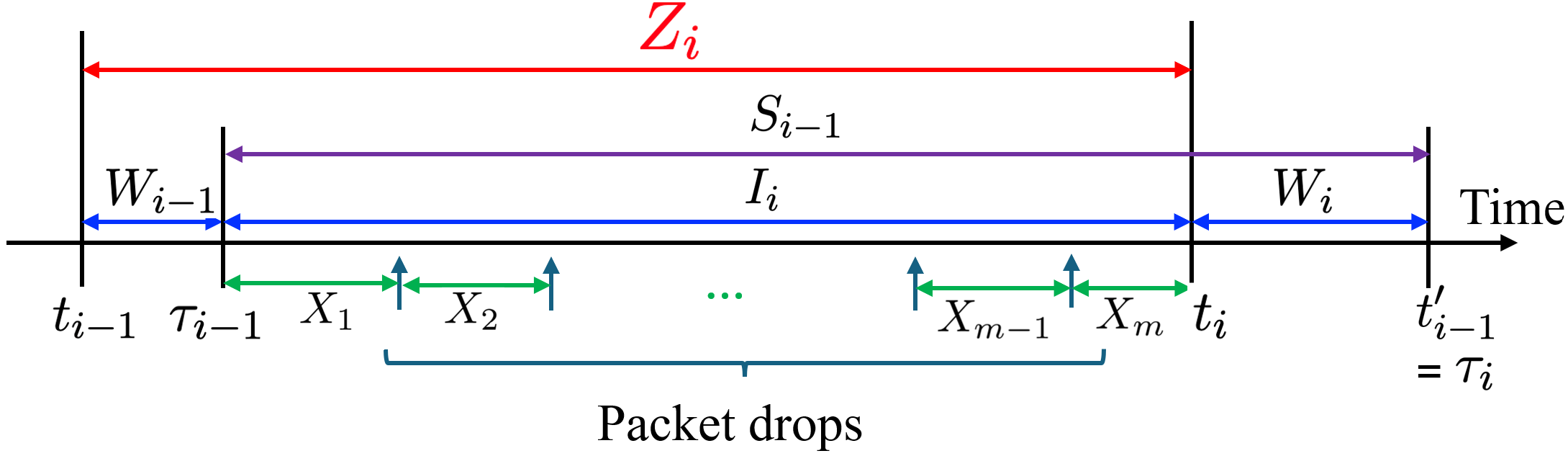}
        \caption{Arrivals occur during the service time of packet $i_{k-1}$.}
        \label{subfig:replacment_time_line2}
    \end{subfigure}
    \caption{Sample path of packet arrivals and departures under the Keep-Fresh policy.}
    \label{fig:replacment_time_line}
\end{figure}

Under the Keep-Fresh policy, the inter-arrival time $Z_i$ consists of the waiting time $W_{i-1}$ and the interval $I_i$  between the service start of the previous packet and the arrival of the next one as shown in Fig.~\ref{fig:replacment_time_line}. The second moment of $Z_i$ is given by
\begin{equation}
    \mathbb{E}[Z_i^2] = \mathbb{E}[W_{i-1}^2] + 2\mathbb{E}[W_{i-1}]\mathbb{E}[I_i] + \mathbb{E}[I_i^2], \label{eq:EZ2}
\end{equation}
where $W_{i-1}$ and $I_i$ are independent due to the memoryless property of inter-arrival and service times.

% \subsubsection{The Waiting Time $W_{i-1}$}

Unlike Keep-Old, where packets arriving at a full buffer are always dropped, Keep-Fresh allows replacement. This creates new dynamics in queueing behavior, particularly in how waiting times are distributed. Using the no-arrival probability during service and the memoryless property of exponential distributions, we derive the expected values
\begin{gather}
    \textstyle\mathbb{E}[W_{i-1}] = \frac{\lambda}{(\lambda+\mu)^2}, \quad \mathbb{E}[W_{i-1}^2] = \frac{2\lambda}{(\lambda+\mu)^3}, \label{eq:EWEW2} \\
    \textstyle\mathbb{E}[I_i] =  \frac{\mu(2\lambda+\mu)}{\lambda(\lambda+\mu)^2} + \frac{\lambda}{\mu(\lambda+\mu)},  \label{eq:EI_total} \\
    \textstyle\mathbb{E}[I^2_i] = \frac{2\mu(3\lambda^2+3\lambda\mu+\mu^2)}{\lambda^2(\lambda+\mu)^3}+\frac{2\lambda}{\mu^2(\lambda+\mu)}. \label{eq:EI2_total}
\end{gather}

The detailed proof is in Appendix~\ref{appendix:Keep_fresh_rec_error}. Combining (\ref{eq:EZ2}), (\ref{eq:EWEW2}), (\ref{eq:EI_total}) and (\ref{eq:EI2_total}), we present the following lemma.
\begin{lemma} \label{lemma:Keep_fresh_rec_error} \it
    The average reconstruction error under the Keep-Fresh policy in an M/M/1/2 queuing system is given by 
    \begin{equation}
        \overline{RE}_{\text{Keep-Fresh}} = \frac{\lambda_{\text{eff}}\mathbb{E}[Z_i^2]}{6},
    \end{equation}
    where $\lambda_{\text{eff}} = \frac{\lambda(\mu^2+\lambda\mu)}{\mu^2 + \lambda\mu+\lambda^2}$ and $\mathbb{E}[Z_i^2] = \frac{2(\lambda^2+\mu^2)}{\lambda^2\mu^2}-\frac{2\lambda(2\lambda+\mu)}{(\lambda+\mu)^4}$.
\end{lemma}
\vspace{0.1in}

From this lemma, we can observe that the average reconstruction error asymptotically approaches $\frac{1}{3\mu}$.
    % \begin{equation}
    %     \textstyle\overline{RE}_{\text{Keep-Fresh}}\rightarrow \frac{1}{3\mu} ~\text{ as } \lambda\rightarrow\infty.
    % \end{equation}
\textit{Thus, despite achieving superior age performance, the Keep-Fresh policy also provides better reconstruction error in moderate-traffic scenarios. However, in heavy-traffic conditions ($\lambda\rightarrow\infty$), both policies achieve the same asymptotic reconstruction error.}

\subsection{Inter-arrival-Aware Dropping Policy} ~\label{subsec:single_IaA}

The Keep-Fresh policy minimizes age but can degrade reconstruction accuracy by discarding historically valuable packets. To address this, we propose an Inter-arrival-Aware dropping policy, which dynamically balances freshness and historical importance based on packet generation times.

Given the generation times of the in-service packet ($t_s$), buffered packet ($t_b$), and new arrival ($t_n$), the policy evaluates temporal gaps and follows the decision rule:
\begin{equation}  
    \begin{split}
        (d_1(t_{\text{n}}),d_2(t_{\text{n}}))=
        \begin{cases}
            (1,0),  &\text{if } t_{\text{b}}- t_{\text{s}} < t_{\text{n}}-t_{\text{b}} \\
            (0,1), &\text{otherwise.}
        \end{cases}
    \end{split} \label{eq:replacing_rule}
\end{equation}
This prioritizes fresher packets when the buffered packet is outdated while preserving historical information if the new arrival adds little value.

Direct analysis of the Inter‐arrival‐Aware dropping policy is intractable due to the continuous-valued generation times, which result in a high‐dimensional Markov chain. To gain insight, we consider the system under heavy traffic ($\lambda\to\infty$). In this regime, we normalize time within a service interval $[\tau_i,\tau_i+S]$, where $S\sim\exp(\mu)$, so that the interval becomes $[0,1]$. By the order‐statistics property of Poisson arrivals, conditioned on there being $k$ arrivals during the service period, their epochs (after scaling by $S$) are distributed as the order statistics of $k$ i.i.d. $\mathrm{Uniform}[0,1]$ random variables.

Under this normalization, in the heavy‐traffic regime, given that the in‐service packet has a waiting time $W_{i-1}$, the waiting time $W_i$ of the buffered packet is well approximated by
\begin{equation}
    W_i = 1 - 2^{\alpha} W_{i-1},
\end{equation}
where
\begin{equation}
    \alpha = \argmin_{\alpha\in\{0,1,2,\ldots\}} \{2^\alpha W_{i-1} > 0.5\}.
\end{equation}
(See Appendix~\ref{appendix:IaA_age} for the detailed proof of existence and uniqueness of the invariant distribution for this recurrence.) Numerical experiments indicate that the invariant distribution of $W$ on $(0,1)$ has mean approximately $0.375$; consequently, the average waiting time of the buffered packet in real time is approximately $0.375S$. Using this result, we present the following lemma.

\begin{lemma} \label{lemma:IaA_age} \it
    The long-term average peak age $\Bar{A}_{\text{IaA}}$ for an $M/M/1/2$ queueing system under the Inter-arrival-Aware dropping policy approximately approach as follows 
    \begin{equation}
        \begin{split}
             \bar{A}_{\text{IaA}} \rightarrow \frac{2.375}{\mu} \text{ as } \lambda\rightarrow\infty.
        \end{split}       
    \end{equation}  
\end{lemma}
Moreover, numerical evaluation suggests that the reconstruction error under the Inter-arrival-Aware dropping policy asymptotically approaches 
\begin{equation} 
    \overline{RE}_{\text{IaA}} \approx \frac{1}{0.362\mu} \quad \text{as} \quad \lambda \to \infty. 
\end{equation} 
Note that this result is obtained numerically without a formal mathematical proof.

Comparing the asymptotic behaviors of three dropping policies, we can see that (1) the Keep-Old policy performs worst in terms of both age and reconstruction error, (2) the Keep-Fresh policy outperforms others in terms of age performance and (3) the Inter-arrival-Aware dropping policy outperforms others in terms of reconstruction error.

\subsection{Threshold-based Inter-arrival-Aware Dropping Policy} \label{subsec:th_IaA}
The Inter-arrival-Aware dropping policy introduced in Section~\ref{subsec:single_IaA} improves reconstruction accuracy by selectively discarding packets based on their inter-arrival times. However, this improvement in reconstruction accuracy comes at the cost of increased AoI compared to the Keep-Fresh policy. In this section, we introduce a threshold-based IaA dropping policy, which introduces a tunable threshold parameter to dynamically balance between AoI and reconstruction error based on their relative importance in the system.

Let $\epsilon$ denote the threshold that determines the decision rule for packet retention. The threshold-based IaA policy is defined as follows:
\begin{equation}  
    \begin{split}
        (d_1(t_{\text{n}}),d_2(t_{\text{n}}))=
        \begin{cases}
            (1,0),  &\text{if } t_{\text{b}}- t_{\text{s}} < t_{\text{n}}-t_{\text{b}}+\epsilon \\
            (0,1), &\text{otherwise.}
        \end{cases}
    \end{split} \label{eq:replacing_rule_threshold}
\end{equation}
The key mechanism of this policy is the threshold $\epsilon$, which controls the trade-off between selecting fresher packets and keeping older packets. When $\epsilon$ is large, the policy tends to accept fresher packets, making its behavior closer to the Keep-Fresh policy. As $\epsilon$ decreases, the policy retains older packets more frequently, mimicking the Keep-Old policy. When $\epsilon = 0$, the decision rule reduces to the Inter-arrival-Aware dropping policy introduced in Section~\ref{subsec:single_IaA}. 

As observed in Sections~\ref{subsec:single_keep_old} and~\ref{subsec:MM12_keep_fresh}, the Keep-Old policy generally performs worse than the Keep-Fresh policy in both age and reconstruction error. This suggests that selecting $\epsilon > 0$ is a reasonable choice to avoid excessive retention of outdated packets. However, analyzing the performance of this policy in a closed-form manner is intractable due to the complex interplay between packet arrivals and update decisions. Instead, in Section~\ref{section:simulation}, numerical results demonstrate how different choices of $\epsilon$ influence the trade-off between age and reconstruction accuracy.

\section{$M/M/1/B+1$ Queueing Systems}

We extend the analysis of the $M/M/1/2$ system to a general $M/M/1/B+1$ system with a buffer capacity of $B$. The server is non-preemptive and follows an LCFS discipline. Unlike in $M/M/1/2$, where all delivered packets are fresh, for $B>2$, some delivered packets may be stale, leading to $i_k \neq k$, where $i_k$ is the index of the $k^{th}$ fresh packet.

\subsection{Keep-Old Policy}

From $(\ref{eq:peak_age_expression})$, the average peak age can be expressed as $\Bar{A} = \mathbb{E}[A_k] = \mathbb{E}[T_{i_{k-1}}] + \mathbb{E}[Y_k]$. The expected system time is given by $\mathbb{E}[T_{i_{k-1}}] = \mathbb{E}[T_{i_k}] = \mathbb{E}[W_{i_k}]+\mathbb{E}[S_{i_k}]$, with $\mathbb{E}[S_{i_k}] = \frac{1}{\mu}$ since service times are  exponentially distributed with rate $\mu$.

The expected waiting time $\mathbb{E}[W_{i_k}]$ depends on the system state at arrival. If the system is empty, the packet is served immediately. If the buffer is full, the waiting time corresponds to the remaining service time of the in-service packet, which follows an exponential distribution with mean $1/\mu$. When the system contains $1 \leq n \leq B-1$ packets, a newly arriving packet remains fresh only if no further arrivals occur before the in-service packet departs, which happens with probability $\frac{\mu}{\lambda+\mu}$. Accounting for these cases, we obtain:
\begin{align}
     \mathbb{E}[W_{i_k}] = \frac{\pi_B}{\mu} + \sum_{n=1}^{B-1} \frac{\pi_n \mu}{\lambda+\mu} \frac{1}{\lambda+\mu}. \label{eq:MM1B_keep_old_EW}
\end{align}

To compute the expected inter-departure time $\mathbb{E}[Y_k]$, we analyze the probability of different arrival scenarios, considering both arrivals that occur during an in-service packet and buffer occupancy states. The detailed proof is in Appendix~\ref{appendix:MM1B_keep_old_age}, leading to
\begin{align}
        &\textstyle\mathbb{E}[Y_k]  = \pi_0\cdot\frac{\lambda^2 + \lambda\mu+\mu^2}{\lambda\mu(\lambda+\mu)} + \left(\pi_B + \frac{\mu}{\lambda+\mu}\sum_{n=1}^{B-1}\pi_n\right)  \label{eq:MM1B_keep_old_ET}  \\
        &\textstyle\hspace{1.3cm}\cdot\left[\sum_{j=0}^{n-1} \left(\frac{\mu}{\lambda+\mu}\right)^{j} \frac{\lambda}{\lambda+\mu} \frac{j+1}{\mu} + \left(\frac{\mu}{\lambda+\mu}\right)^{n} \left(\frac{n}{\mu}+\frac{1}{\lambda}\right)\right]. \nonumber
\end{align}

Combining and rearranging $\mathbb{E}[S_k] = \frac{1}{\mu}$, (\ref{eq:MM1B_keep_old_EW}), (\ref{eq:MM1B_keep_old_ET}) and (\ref{eq:steady_prob}), we present the following lemma.
\begin{lemma}\label{lemma:MM1B_keep_old_age} \it
    The long-term average peak age $\Bar{A}_{\text{Keep-Old}}(B)$ for an M/M/1/B+1 queueing system under the Keep-Old policy with a LCFS discipline are given by
    \begin{equation}
        \begin{split}
            &\Bar{A}_{\text{Keep-Old}}(B) \textstyle= \frac{1}{\mu} + \frac{1}{\mathcal{C}_1} \Big[\frac{1}{\lambda}+\frac{1}{\mu-\lambda} \left(1+\frac{\lambda\mu}{(\lambda+\mu)^2}\right) \\
            &\hspace{2.cm}\textstyle+\left(\frac{2}{\mu}-\frac{1}{\mu-\lambda}\left(1+\frac{\mu^2}{(\lambda+\mu)^2}\right)\right)\rho^B \Big], 
        \end{split}
    \end{equation}
    where $\mathcal{C}_1 = 1+\frac{\lambda\mu}{\mu^2-\lambda^2} - \frac{\lambda^2}{\mu^2-\lambda^2}\rho^B$ and  $\rho = \frac{\lambda}{\mu}$. 
\end{lemma}
\vspace{0.1in}

From this lemma, we can observe the long-term average peak age under the Keep-Old policy exhibits specific asymptotic behaviors. As $\lambda \rightarrow \infty$ with a fixed buffer size $B \ge 1$, the average peak age approaches $\frac{3}{\mu}$. For given $\lambda \ge \mu$, as $B\rightarrow \infty$, it approaches $\frac{3}{\mu} + \frac{1}{\lambda+\mu}$. In the low-traffic regime  ($\lambda < \mu$), as $B\rightarrow \infty$, the peak age converges to $\frac{1}{\mu} + \frac{\mu(2\lambda^2 + 2\lambda\mu + \mu^2)}{\lambda(\lambda+\mu)(\mu^2+\lambda\mu-\lambda^2)}$. 

For reconstruction error, we analyze the second moment $\mathbb{E}[Z_{i}^2]$ of the inter-arrival time, considering cases where a fresh packet arrives when the buffer is full or not. Using probability-weighted expectations, we have 
% For the reconstruction error, we express it in terms of the second moment of inter-arrival times. If a fresh packet arrives when the system is not full, the inter-arrival time follows an exponential distribution with mean $1/\lambda$. However, if the system is full at arrival, the inter-arrival time consists of both the remaining service time and the time until the next arrival after service completion. Since these two components are independent, we compute their second moments separately and combine them to obtain the final reconstruction error expression as
\begin{equation}
     \mathbb{E}[Z_{\text{Keep-Old}}^2] = 2\pi_B \left(\frac{1}{\mu^2}+\frac{1}{\lambda\mu}+\frac{1}{\lambda^2}\right) + \frac{2}{\lambda^2} \sum_{n=0}^{B-1} \pi_n.
\end{equation}
The full derivation is in Appendix~\ref{appendix:MM1B_keep_old_error}. 
Rearranging this with (\ref{eq:steady_prob}), we present the following lemma.
\begin{lemma} \label{lemma:MM1B_keep_old_error}\it
    The long-term average reconstruction error $\overline{RE}_{\text{K-O}}(B)$ for an M/M/1/B+1 queueing system under the Keep-Old policy with a LCFS discipline are given by
    \begin{equation}
        \overline{RE}_{\text{Keep-Old}}(B) = \frac{\lambda_{\text{eff}}\mathbb{E}[Z_{\text{K-O}}^2]}{6},
    \end{equation}
    where $\lambda_{\text{eff}} = \frac{\lambda (1-\rho^{B+1})}{1-\rho^{B+2}}$, $\mathbb{E}[Z_{\text{K-O}}^2] = \frac{2}{\mathcal{C}_2(\mu-\lambda)} \left(\frac{\mu}{\lambda^2}-\frac{\lambda}{\mu^2} \rho^B\right)$
    % \begin{equation*}
    %  \lambda_{\text{eff}} = \frac{\lambda (1-\rho^{B+1})}{1-\rho^{B+2}}, ~\mathbb{E}[Z_{\text{K-O}}^2] = \frac{2}{\mathcal{C}_2(\mu-\lambda)} \left(\frac{\mu}{\lambda^2}-\frac{\lambda}{\mu^2} \rho^B\right) 
    % \end{equation*}
    and $\mathcal{C}_2 = \frac{\mu}{\mu-\lambda}-\frac{\lambda}{\mu-\lambda}\rho^B$.
\end{lemma}
\vspace{0.1in}

For low traffic ($\lambda < \mu$), reconstruction error approaches $\frac{1}{3\lambda}$ as $B \to \infty$. In high traffic ($\lambda > \mu$), it converges to $1/(3\mu)$, showing service rate dominance. For fixed $B$, as $\lambda \to \infty$, it stabilizes at $\frac{1}{3\mu}$.

\subsection{Keep-Fresh Policy}
\subsubsection{The Average Peak Age}

The expected service time $\mathbb{E}[S_{i_k}] = \frac{1}{\mu}$ remains policy-independent. However, under the Keep-Fresh policy, a packet remains fresh only if no new arrivals occur during the in-service packet’s remaining service time, which occurs with probability $\frac{\mu}{\lambda+\mu}$. Then, the expected waiting time is given by
% The expected service time $\mathbb{E}[S_{i_k}]$ is independent of policies, which is $\frac{1}{\mu}$ due to the exponential distribution. Under the LCFS Keep-Fresh policy, a packet remains fresh if no new arrivals occur during the remaining service time of the in-service packet, which occurs with probability $\frac{\mu}{\lambda+\mu}$. Given this, the expected waiting time is given by
\begin{equation}
    \begin{split}
        \mathbb{E}[W_{i_k}] = \sum_{n=1}^{B+1} \frac{\mu\pi_n}{\lambda+\mu} \frac{1}{\lambda+\mu}.  \label{eq:MM1B_keep_fresh_EW}
    \end{split}
\end{equation}

\begin{figure*}[!ht]
    \begin{equation}
    \begin{split}
        \mathbb{E}[Y_k] & = \pi_0 \left(\frac{\lambda}{(\lambda+\mu)\mu} + \frac{1}{\lambda}\right) + \frac{\mu}{\lambda+\mu}\sum_{n=1}^{B-1}\pi_n \left(\sum_{j=0}^{n-1} \left(\frac{\mu}{\lambda+\mu}\right)^{j} \frac{\lambda}{\lambda+\mu} \frac{j+1}{\mu} + \left(\frac{\mu}{\lambda+\mu}\right)^{n} \left(\frac{n}{\mu}+\frac{1}{\lambda}\right)\right) \\
        &\hspace{1cm}+ \frac{\mu}{\lambda+\mu}(\pi_B+\pi_{B+1})\left(\sum_{j=0}^{B-1} \left(\frac{\mu}{\lambda+\mu}\right)^{j} \frac{\lambda}{\lambda+\mu} \frac{j+1}{\mu} + \left(\frac{\mu}{\lambda+\mu}\right)^{B} \left(\frac{B}{\mu}+\frac{1}{\lambda}\right)\right).
        \label{eq:MM1B_keep_fresh_ET} 
    \end{split}
    \end{equation}
    \hrule
\end{figure*}

For the inter-departure time $\mathbb{E}[Y_k]$, the expectation remains the same as in the Keep-Old policy when the system is empty, but differs when the buffer is non-empty due to different queueing dynamics. However, a similar analytical approach can be applied with slight modifications. The expected inter-departure time is given by (\ref{eq:MM1B_keep_fresh_ET}), with detailed derivations provided in Appendix~\ref{appendix:MM1B_keep_fresh_age}.
%
% For the inter-departure time $\mathbb{E}[Y_k]$, the expectation under an empty buffer matches the Keep-Old policy. Under the Keep-Fresh policy, if a packet arrives when the buffer is not empty, it remains fresh with probability $\frac{\mu}{\lambda+\mu}$. Unlike the Keep-Old policy, packets arriving when the buffer is full may replace the tail packet and still risk being dropped by subsequent arrivals. However, for packets that do remain fresh, the expected inter-departure time remains the same as in the Keep-Old policy due to the LCFS structure. Using the reasoning and results from the Keep-Old policy, we obtain the expected inter-departure time under the Keep-Fresh policy as (\ref{eq:MM1B_keep_fresh_ET}).
%
%
%
%
Combining and rearranging $\mathbb{E}[S_{i_k}]=\frac{1}{\mu}$, (\ref{eq:MM1B_keep_fresh_EW}), (\ref{eq:MM1B_keep_fresh_ET}) and (\ref{eq:steady_prob}), we present the following lemma.
\begin{lemma} \label{lemma:MM1B_keep_fresh_age}\it
    The long-term average peak age $\Bar{A}_{\text{Keep-Fresh}}(B)$ for an M/M/1/B+1 queueing system  under the Keep-Fresh policy with a LCFS discipline are given by
    \begin{align}
         \Bar{A}_{\text{Keep-Fresh}}(B) &\textstyle= \frac{1}{\mu} + \frac{1}{\mathcal{C}_1} \bigg[\frac{1}{\lambda}+\frac{1}{\mu-\lambda} \left(1+\frac{\lambda\mu}{(\lambda+\mu)^2}\right) \nonumber\\
         &\textstyle+ \left(\frac{1}{\mu}-\frac{1}{\mu-\lambda}\left(1+\frac{\lambda^2}{(\lambda+\mu)^2}\right)\right)\rho^B\bigg],
    \end{align}
    where $\mathcal{C}_1 = 1+\frac{\lambda\mu}{\mu^2-\lambda^2} - \frac{\lambda^2}{\mu^2-\lambda^2}\rho^B$ and  $\rho = \frac{\lambda}{\mu}$.
\end{lemma}
\vspace{0.1in}

As $\lambda \rightarrow \infty$ for a given $B \geq 1$, the average peak age under the Keep-Fresh policy approaches $\frac{2}{\mu}$. For $\lambda \geq \mu$, as $B \rightarrow \infty$, it converges to $\frac{2}{\mu} + \frac{1}{\lambda} + \frac{1}{\lambda+\mu}$. When $\lambda < \mu$, $\Bar{A}_{\text{Keep-Fresh}}(B)$ matches the Keep-Old policy, indicating that under low traffic, both policies achieve the same asymptotic peak age as $B$ grows. However, for $\lambda \geq \mu$, Keep-Fresh maintains an advantage by prioritizing fresher updates.

% Note that  $\Bar{A}_{\text{Keep-Fresh}}(B)\rightarrow \frac{2}{\mu}$ as $\lambda \rightarrow \infty$ for a given $B \ge 1$. Further, for a given $\lambda \ge \mu$,  $\Bar{A}_{\text{Keep-Fresh}}(B)\rightarrow \frac{2}{\mu} + \frac{1}{\lambda} + \frac{1}{\lambda+\mu}$ as $B\rightarrow \infty$. For the case $\lambda < \mu$, $\Bar{A}_{\text{Keep-Fresh}}(B)\rightarrow \frac{1}{\mu} + \frac{\mu(2\lambda^2 + 2\lambda\mu + \mu^2)}{\lambda(\lambda+\mu)(\mu^2+\lambda\mu-\lambda^2)}$ as $B\rightarrow\infty$, which matches the behavior of the Keep-Old policy. The equivalence between the two policies holds only for the case where $\lambda < \mu$. This reflects that, under low traffic conditions, both policies achieve the same asymptotic average peak age in large-buffer scenarios. However, when $\lambda \geq \mu$, the Keep-Fresh policy shows a distinct advantage due to its prioritization of fresher updates.

For reconstruction error, we analyze the second moment $\mathbb{E}[Z_i^2]$ of inter-arrival times by considering cases where the buffer is full or not. When not full, inter-arrival time follows an exponential distribution with mean $\frac{1}{\lambda}$. When full, it depends on whether a new packet arrives before the in-service packet departs, incorporating both the remaining service time and the next arrival time. The detailed derivation is in Appendix~\ref{appendix:MM1B_keep_fresh_error}.
% Packets arriving when the buffer is not full are always served, and their inter-arrival times remain unaffected by the policy. In this case, the second moment of the inter-arrival time follows the standard result, given by $\mathbb{E}[Z_i^2] = \frac{2}{\lambda^2}$ for arrivals occurring when the buffer is not full.
% 
% When the buffer contains $B$ packets, the inter-arrival time depends on whether new arrivals occur during the remaining service time of the in-service packet. When no new arrivals occur during the service time, the inter-arrival time is simply the sum of independent exponentially distributed waiting, service, and arrival intervals, leading to a straightforward summation of their variances. When new arrivals occur, the inter-arrival time depends on the interval between the departure of the in-service packet and the latest arrival, which follows a recursive structure based on the arrival process. These cases are analyzed separately, and their contributions are weighted by their respective probabilities, leading to the final expression for $\mathbb{E}[Z_{\text{Keep-Fresh}}^2]$. The detailed derivations are provided in Appendix~\RED{F}, and we present the following lemma.
\begin{lemma} \label{lemma:MM1B_keep_fresh_error}\it
    The long-term average reconstruction error $\overline{RE}_{\text{Keep-Fresh}}(B)$ for an M/M/1/B+1 queueing system under the Keep-Fresh policy with a LCFS discipline are given by
    \begin{equation}
        \overline{RE}_{\text{Keep-Fresh}}(B) = \frac{\lambda_{\text{eff}}\mathbb{E}[Z_{\text{K-F}}^2]}{6},
    \end{equation}
    where $ \lambda_{\text{eff}} = \frac{\lambda (1-\rho^{B+1})}{1-\rho^{B+2}}$ and 
    % \begin{equation}
    % \lambda_{\text{eff}} = \frac{\lambda (1-\rho^{B+1})}{1-\rho^{B+2}}, \text{ and }\nonumber
    % \end{equation}
    \begin{equation}
        \textstyle\mathbb{E}[Z_{\text{K-F}}^2] = \frac{1}{\mathcal{C}_2}\left( \frac{2\mu}{(\mu-\lambda)\lambda^2}+ \left(\mathcal{Z}_1 + \mathcal{Z}_2 - \frac{2\mu^2}{(\mu-\lambda)\lambda^3}\right)\rho^B\right), \nonumber
    \end{equation}
    where 
    \begin{align}  
        \mathcal{Z}_1 &\textstyle= \frac{2\mu^3(6\lambda^2+4\lambda\mu+\mu^2)+2\lambda^3(\lambda^2+3\lambda\mu+3\mu^2)}{\lambda^2 \mu^2 (\lambda+\mu)^3}, \nonumber \\
        \mathcal{Z}_2 &\textstyle= \frac{2\mu^2(3\lambda^2+3\lambda\mu+\mu^2)}{\lambda^3(\lambda+\mu)^3} + \frac{2}{\mu(\lambda+\mu)} \text{ and } \mathcal{C}_2 = \frac{\mu}{\mu-\lambda}-\frac{\lambda}{\mu-\lambda}\rho^B.\nonumber
    \end{align}
    % and $\mathcal{C}_2 = \frac{\mu}{\mu-\lambda}-\frac{\lambda}{\mu-\lambda}\rho^B$.
\end{lemma}
\vspace{0.1in}

For $\lambda < \mu$, the reconstruction error asymptotically matches the Keep-Old policy, approaching $\frac{1}{3\lambda}$ as $B\rightarrow\infty$. When $\lambda > \mu$, it converges to a lower value than the Keep-Old policy, given by $\frac{\lambda-\mu}{\lambda}(\mathcal{Z}_1+\mathcal{Z}_2 + \frac{2\mu^2}{(\lambda-\mu)\lambda^3})$. For fixed $B$, as $\lambda\rightarrow\infty$, it approaches $\frac{1}{3\mu}$, again aligning with the Keep-Old policy.

\subsection{Threshold-based Inter-arrival-Aware Dropping Policy}

The Inter-arrival-Aware dropping policy dynamically evaluates inter-arrival time gaps to determine which packet should be dropped when the buffer is full, aiming to preserve packets that maximize temporal diversity. It considers the timing of all packets in the system, including those in service, buffered, and newly arriving, ensuring an informed dropping decision.

% The Inter-arrival-Aware dropping policy for the $M/M/1/B+1$ queueing system dynamically evaluates inter-arrival time gaps between packets to determine which packet should be dropped when the buffer reaches capacity. By considering the timing of all packets in the system, this policy aims to preserve those that contribute most to temporal diversity. The inter-arrival time gaps account for the packet in service, the buffered packets, and the newly arriving packet, allowing the system to assess the distribution of packet arrival times before making a dropping decision.

Let $\mathcal{S}$, $\mathcal{B}$ and $\mathcal{D}$ denote packets in service, the buffer, and delivered, respectively. The generation time of the $i^{th}$ buffered packet is denoted as $t_{\text{b}_i}$ for $i=1,...,B$, and $t_{\text{b}_{B+1}}$ denotes the newly arriving packet. For each packet $i$, the most recent preceding packet is indexed as $\text{b}_i^* = \max\{j < i : j\in \mathcal{S}\cup\mathcal{B}\cup\mathcal{D}\}$. Then, the inter-arrival gap is given by $\Delta_i = t_{\text{b}_i} - t_{\text{b}_i^*}$ for $i = 1,...,B$, and $\Delta_{B+1} = t_{b_{B+1}}-t_{b_{B+1}^*} + \epsilon$ for the new packet.

When the buffer is full, the packet with the smallest inter-arrival gap is dropped, ensuring diverse temporal representation. The dropping decision is given by 
\begin{equation}
    j^* = \argmin_{j\in\{1,...,B+1\}} \Delta_{j}.
\end{equation}
Thus, $d_{j^*}(t_{\text{b}_{B+1}}) = 1$, while $d_j(t_{\text{b}_{B+1}}) = 0$ for $j \neq j^*$.

A theoretical analysis of this policy is challenging due to interdependent dropping decisions and dynamic arrivals. Instead, we evaluate its performance numerically in Section~\ref{section:simulation}, examining its impact under varying traffic conditions.

% This procedure is visually illustrated in Fig.~\ref{fig:MM1B_dropping_policy}, which depicts the calculation of inter-arrival gaps and the decision-making process for packet dropping. 

% Let the generation times of the packets in the system be $t_{\text{s}},t_{\text{b}_1},...,t_{\text{b}_B}$ and $t_{\text{n}}$. The inter-arrival time gaps are defined as:
% \begin{equation}
%     \begin{split}
%         &\Delta_{1,0} := t_{\text{b}_1}-t_{\text{s}}, \\
%         &\Delta_{j,j-1} := t_{\text{b}_j}-t_{\text{b}_{j-1}}, ~\text{ for } j=2,...,B,\\
%         &\Delta_{B+1,B} := t_{\text{n}}-t_{\text{b}_B}.
%     \end{split}
% \end{equation}

% \begin{figure}
%     \centering
%     \includegraphics[width=\linewidth]{Fig_MM1B_dropping_policy.png}
%     \caption{Adaptive dropping policy under an $M/M/1/B+1$ queueing system.}
%     \label{fig:MM1B_dropping_policy}
% \end{figure}

\section{Numerical Results} \label{section:simulation}

In this section, we present numerical results to validate the theoretical analysis and gain further insights into the performance of the proposed policies under various system configurations. The simulations are designed to illustrate key metrics such as the average peak age and the reconstruction error highlighting the impact of buffer size, arrival rates, and service rates.

\subsection{Single-Buffer Queueing System}
In this subsection, we analyze the performance of single-buffer systems, focusing on the $M/M/1/2$ case while also discussing potential extensions to $M/G/1/2$ and $G/M/1/2$ systems.

\subsubsection{$M/M/1/2$ Queue}

\begin{figure}
    \centering
    \begin{subfigure}[b]{0.24\textwidth}
        \centering
        \includegraphics[width=\linewidth]{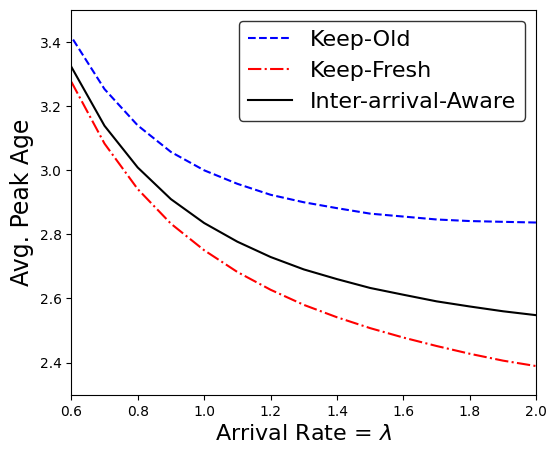}
        \caption{Average peak age.}
        \label{subfig:MM12_peak_age1}
    \end{subfigure}
    \hfill
    \begin{subfigure}[b]{0.24\textwidth}
        \centering
        \includegraphics[width=\linewidth]{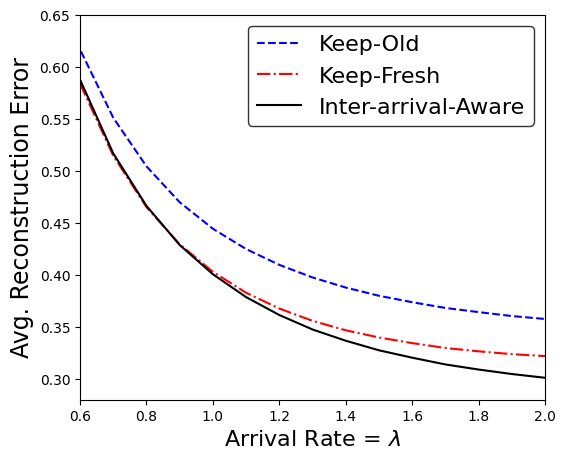}
        \caption{Average reconstruction error.}
        \label{subfig:MM12_rec_error1}
    \end{subfigure}
    \caption{Comparison of three packet-dropping policies in an M/M/1/2 queueing system with service rate is $\mu=1$.}
    \label{fig:MM12_comparison1}
\end{figure}

We first consider an $M/M/1/2$ system with a fixed service rate $\mu = 1$.
%, comparing the Keep-Old, Keep-Fresh, and Inter-arrival-Aware Dropping policies.
Fig.~\ref{fig:MM12_comparison1} shows that the Keep-Fresh policy achieves the lowest average peak age by prioritizing the freshest packets, whereas the Inter-arrival-Aware policy minimizes reconstruction error by dynamically selecting which packets to drop. At $\lambda = 2$, Keep-Fresh reduces peak age by $6.64\%$ compared to Inter-Arrival-Aware, while Inter-Arrival-Aware improves reconstruction accuracy by $6.46\%$ over Keep-Fresh. As $\lambda$ increases, Keep-Fresh maintains its advantage in age performance, whereas Inter-Arrival-Aware continues to achieve superior reconstruction accuracy.

\begin{figure}
    \centering
    \begin{subfigure}[b]{0.24\textwidth}
        \centering
        \includegraphics[width=\linewidth]{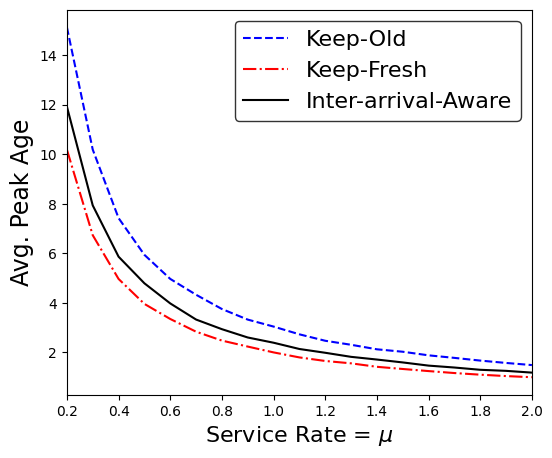}
        \caption{Average peak age.}
        \label{subfig:MM12_peak_age2}
    \end{subfigure}
    \hfill
    \begin{subfigure}[b]{0.24\textwidth}
        \centering
        \includegraphics[width=\linewidth]{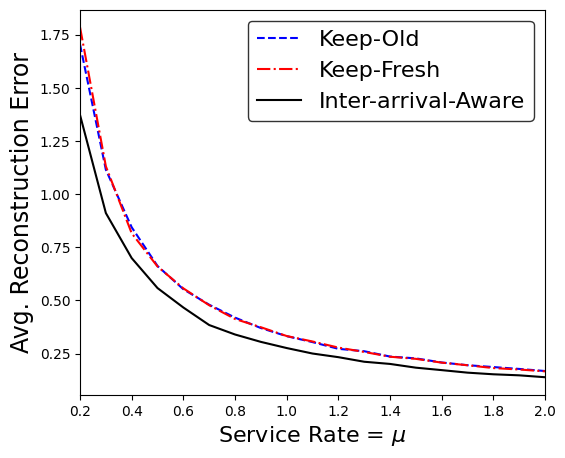}
        \caption{Average reconstruction error.}
        \label{subfig:MM12_rec_error2}
    \end{subfigure}
    \caption{Comparison of three packet-dropping policies in an $M/M/1/2$ queueing system under heavy traffic ($\lambda=10^3$).}
    \label{fig:MM12_comparison2}
\end{figure}

Under heavy traffic conditions ($\lambda = 10^3$), Fig.~\ref{fig:MM12_comparison2} demonstrates that increasing $\mu$ reduces both peak age and reconstruction error across all policies. Faster service enables fresher updates, benefiting Keep-Fresh in terms of age and allowing Inter-arrival-Aware to retain more informative packets for reconstruction. Keep-Fresh achieves a $15.63\%$ lower peak age than Inter-Arrival-Aware, ensuring fresher updates, while Inter-Arrival-Aware reduces reconstruction error by $16.92\%$ compared to Keep-Fresh, effectively preserving historical information for improved reconstruction accuracy.

\begin{figure}
    \centering
    \begin{subfigure}[b]{0.24\textwidth}
        \centering
        \includegraphics[width=\linewidth]{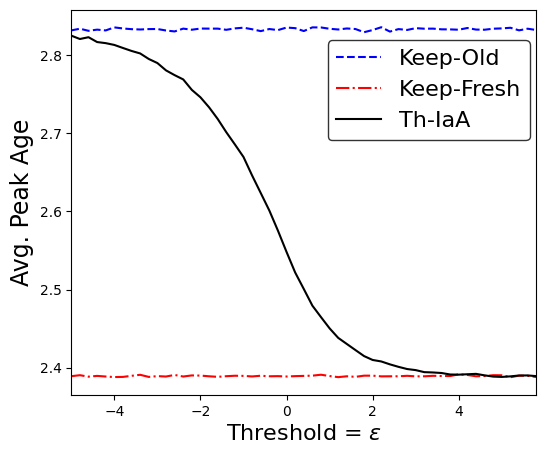}
        \caption{Average peak age.}
        \label{subfig:MM12_peak_age3}
    \end{subfigure}
    \hfill
    \begin{subfigure}[b]{0.24\textwidth}
        \centering
        \includegraphics[width=\linewidth]{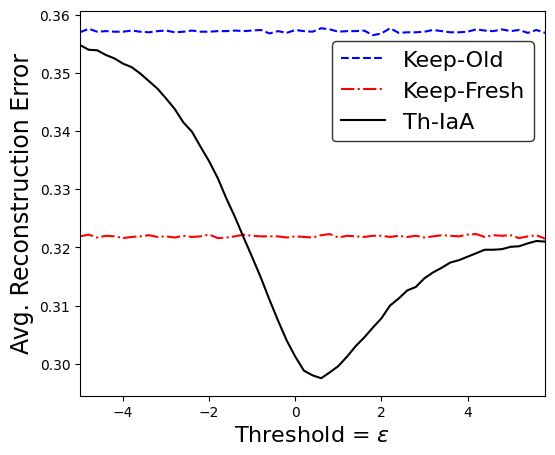}
        \caption{Average reconstruction error.}
        \label{subfig:MM12_rec_error3}
    \end{subfigure}
    \caption{Comparison of three packet-dropping policies in an $M/M/1/2$ queueing system with $\lambda=2$ and $\mu=1$.}
    \label{fig:MM12_comparison3}
\end{figure}

\begin{figure}
    \centering
    \begin{subfigure}[b]{0.24\textwidth}
        \centering
        \includegraphics[width=\linewidth]{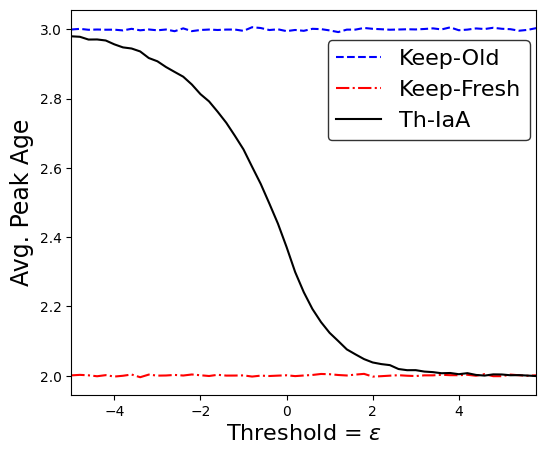}
        \caption{Average peak age.}
        \label{subfig:MM12_peak_age4}
    \end{subfigure}
    \hfill
    \begin{subfigure}[b]{0.24\textwidth}
        \centering
        \includegraphics[width=\linewidth]{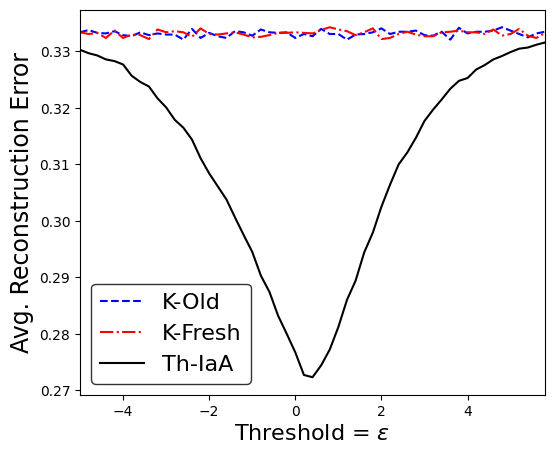}
        \caption{Average reconstruction error.}
        \label{subfig:MM12_rec_error4}
    \end{subfigure}
    \caption{Comparison of three packet-dropping policies in an $M/M/1/2$ queueing system with $\lambda = 10^3$ and $\mu=1$.}
    \label{fig:MM12_comparison4}
\end{figure}

We next examine the threshold-based Inter-arrival-Aware (Th-IaA) policy. As shown in Fig.\ref{fig:MM12_comparison3} and Fig.\ref{fig:MM12_comparison4}, the Th-IaA policy introduces a tunable threshold parameter $\epsilon$ to balance the trade-off between average peak age and reconstruction error more flexibly than the original Inter-arrival-Aware policy.

For $\lambda = 2$ and $\mu = 1$, the choice of $\epsilon \approx 0.6$ minimizes the reconstruction error while slightly reducing the average peak age compared to the original IaA policy ($\epsilon = 0$). Specifically, setting $\epsilon = 0.6$ reduces the average peak age by approximately $2.70\%$ and decreases the reconstruction error by about $1.23\%$. However, compared to the Keep-Fresh policy, the Th-IaA policy with $\epsilon = 0.6$ results in a $14.38\%$ higher average peak age but improves reconstruction accuracy by $16.83\%$. This suggests that a moderate threshold value allows the Th-IaA policy to retain more informative packets for historical reconstruction while partially mitigating the age penalty.

In extremely heavy traffic conditions ($\lambda = 10^3$ and $\mu = 1$), selecting a threshold of  $\epsilon \approx 0.4$ appears to be more effective. With this setting, the average peak age is reduced by $5.50\%$ and the reconstruction error by $1.59\%$ compared to the original IaA policy.  In comparison to the Keep-Fresh policy, the Th-IaA policy with $\epsilon = 0.4$ results in a 
$12.00\%$ higher average peak age but improves reconstruction accuracy by $18.27\%$. These results emphasize the ability to strike a more balanced trade-off between age and reconstruction error.

%%%----------------

\begin{figure}
    \centering
    \begin{subfigure}[b]{0.24\textwidth}
        \centering
        \includegraphics[width=\linewidth]{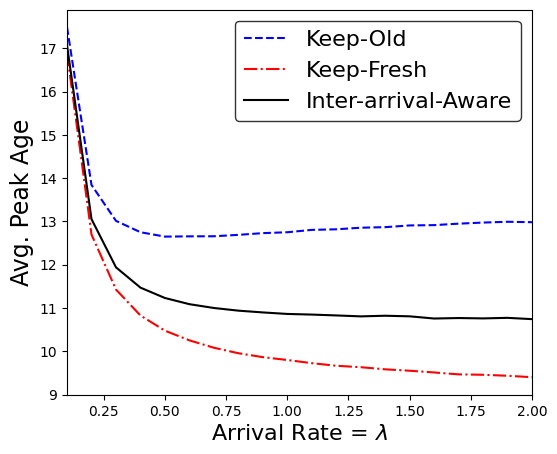}
        \caption{Average peak age.}
        \label{subfig:MG12_peak_age}
    \end{subfigure}
    \hfill
    \begin{subfigure}[b]{0.24\textwidth}
        \centering
        \includegraphics[width=\linewidth]{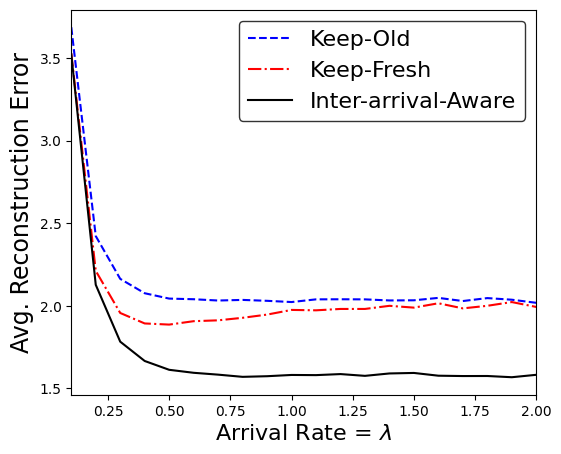}
        \caption{Average reconstruction error.}
        \label{subfig:MG12_rec_error}
    \end{subfigure}
   \caption{$M/G/1/2$ queueing system with log-normal service times with parameters $\mu=1$ and $\sigma=1$. }
    \label{fig:MG12_comparison}
\end{figure}

\begin{figure}
    \centering
    \begin{subfigure}[b]{0.24\textwidth}
        \centering
        \includegraphics[width=\linewidth]{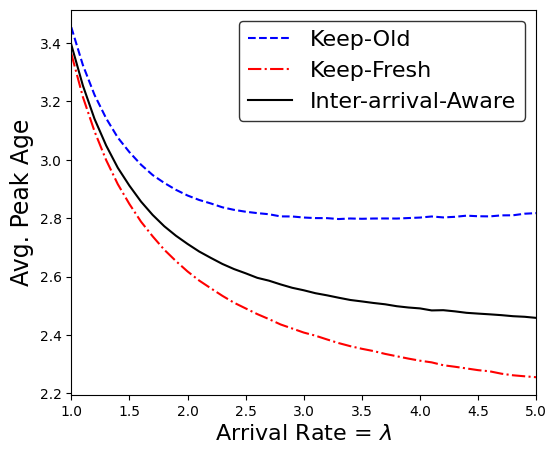}
        \caption{Average peak age.}
        \label{subfig:GM12_peak_age1}
    \end{subfigure}
    \hfill
    \begin{subfigure}[b]{0.24\textwidth}
        \centering
        \includegraphics[width=\linewidth]{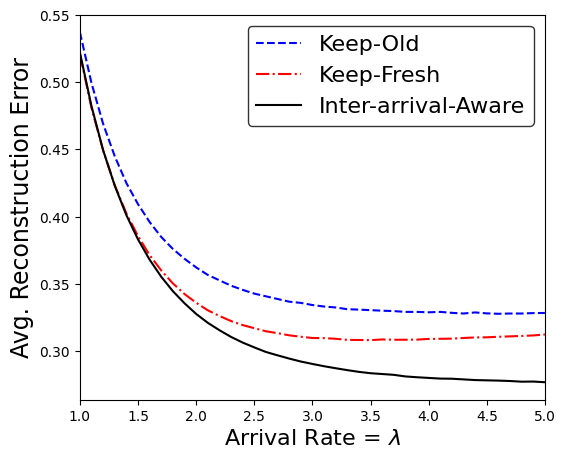}
        \caption{Average reconstruction error.}
        \label{subfig:GM12_rec_error1}
    \end{subfigure}
    \caption{$G/M/1/2$ queueing system with Erlang-2 arrival process and service rate $\mu=1$. }
    \label{fig:GM12_comparison1}
\end{figure}

\begin{figure}
    \centering
    \begin{subfigure}[b]{0.24\textwidth}
        \centering
        \includegraphics[width=\linewidth]{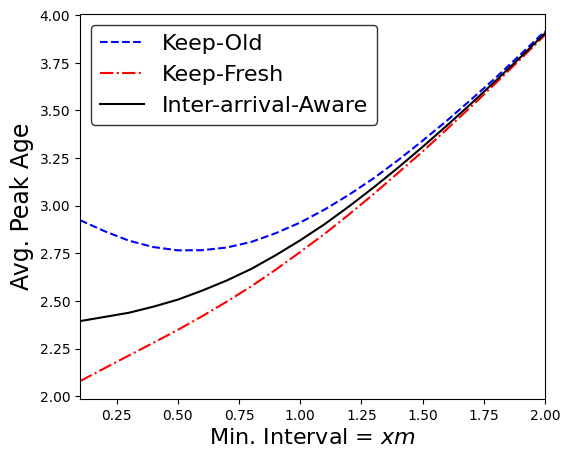}
        \caption{Average peak age.}
        \label{subfig:GM12_peak_age2}
    \end{subfigure}
    \hfill
    \begin{subfigure}[b]{0.24\textwidth}
        \centering
        \includegraphics[width=\linewidth]{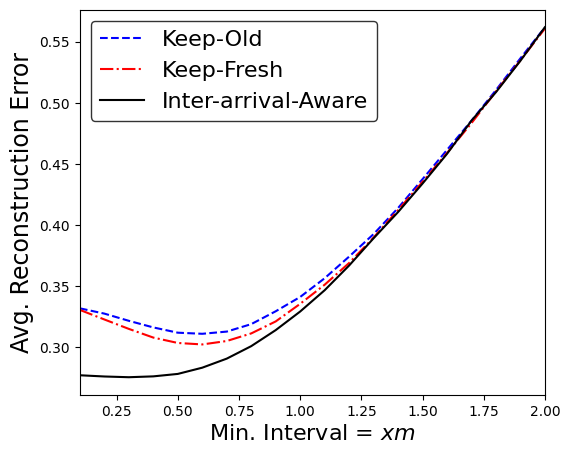}
        \caption{Average reconstruction error.}
        \label{subfig:GM12_rec_error2}
    \end{subfigure}
    \caption{$G/M/1/2$ with Pareto inter-arrival times with $\alpha=3.5$ and varying $x_m$, and service rate $\mu=1$.}
    \label{fig:GM12_comparison2}
\end{figure}

\subsubsection{$M/G/1/2$ Queue}

The $M/G/1/2$ queue extends the analysis by allowing the service times to follow a general distribution. In this study, we consider a log-normal distribution with parameters $\mu = 1$ and $\sigma = 1$, which introduces variability in service times. Fig.~\ref{fig:MG12_comparison} shows the average peak age and reconstruction error for the three packet-dropping policies as the arrival rate varies from $0.2$ to $4$. The results demonstrate similar trends to the $M/M/1/2$. However, the variability in service times slightly widens the performance gap between the policies, with Inter-arrival-Aware maintaining its advantage in reconstruction error and Keep-Old performing best in terms of age.

\subsubsection{$G/M/1/2$ Queue}

For the $G/M/1/2$ queue, the inter-arrival times are allowed to follow general distributions. We first consider the case of $Erlang-2$ arrivals. The structured nature of Erlang arrivals reduces randomness in queueing dynamics. As shown in Fig.~\ref{fig:GM12_comparison1}, the trends in peak age and reconstruction error are consistent with those observed in the $M/M/1/2$ queue. The Keep-Old policy achieves the lowest peak age, while Inter-arrival-Aware minimizes the reconstruction error. We then examine Pareto-distributed inter-arrival times, which introduce bursty traffic with high variability. The heavy-tailed nature of the Pareto distribution results in sporadic bursts of arrivals, creating challenges for all policies. Fig.~\ref{fig:GM12_comparison2} shows that Inter-arrival-Aware handles these bursts most effectively, achieving the lowest reconstruction error, while the Keep-Old policy manages age performance better under such variability.

\subsection{$B$-Buffer Queueing System}
\begin{figure}
    \centering
    \begin{subfigure}[b]{0.24\textwidth}
        \centering
        \includegraphics[width=\linewidth]{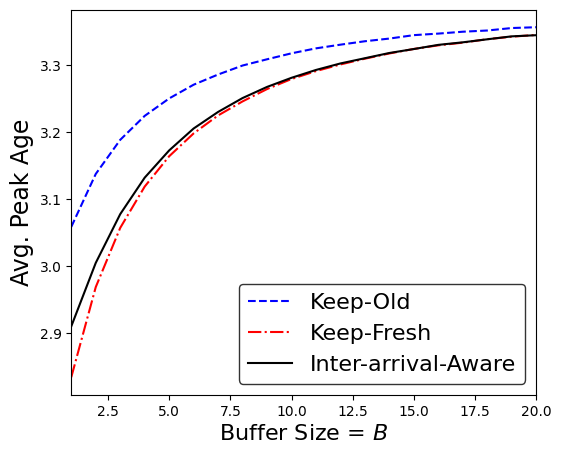}
        \caption{Average peak age.}
        \label{subfig:MM1B_lt_peak_age}
    \end{subfigure}
    \hfill
    \begin{subfigure}[b]{0.24\textwidth}
        \centering
        \includegraphics[width=\linewidth]{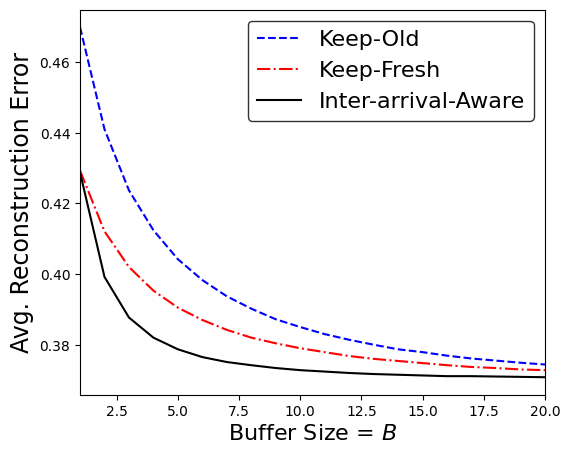}
        \caption{Average reconstruction error.}
        \label{subfig:MM1B_lt_rec_error}
    \end{subfigure}
    \caption{When $\lambda = 0.9$ and $\mu=1$.}
    \label{fig:MM1B_lt_comparison}
\end{figure}

We next examine the effect of buffer size $B$ on system performance in the $M/M/1/B+1$ queueing system under various traffic conditions, where the arrival rate $\lambda$ varies while the service rate remains fixed at $\mu=1$. In a low-traffic scenario with $\lambda = 0.9$ (Fig.~\ref{fig:MM1B_lt_comparison}), packet drops are infrequent, leading to minimal performance differences between policies. As $B$ increases, peak age initially rises since staled packets accumulate in the buffer, increasing the chance of transmitting older packets instead of fresher ones. However, as $B$ grows further, this effect diminishes. Meanwhile, reconstruction error steadily decreases as a larger buffer retains more packets, improving state estimation by reducing information loss.

\begin{figure}
    \centering
    \begin{subfigure}[b]{0.24\textwidth}
        \centering
        \includegraphics[width=\linewidth]{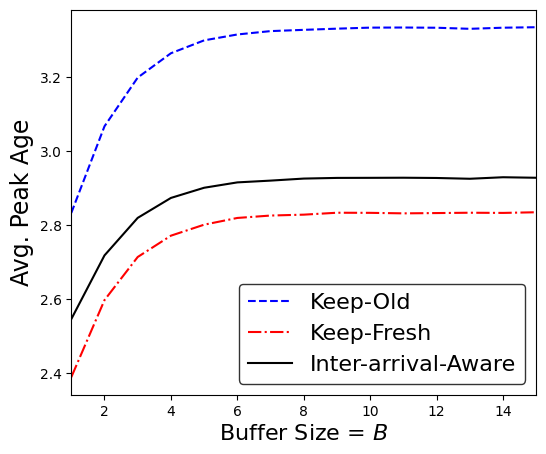}
        \caption{Average peak age.}
        \label{subfig:MM1B_mt_peak_age}
    \end{subfigure}
    \hfill
    \begin{subfigure}[b]{0.24\textwidth}
        \centering
        \includegraphics[width=\linewidth]{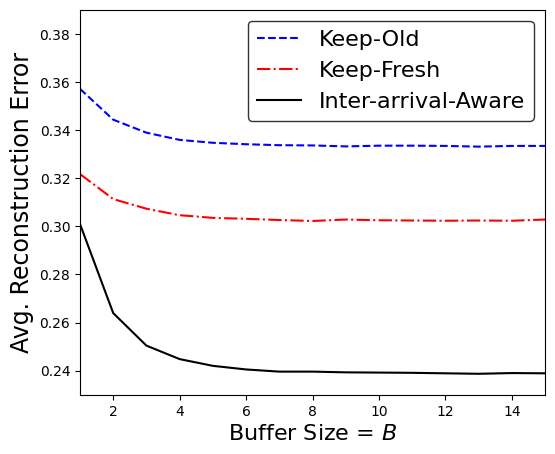}
        \caption{Average reconstruction error.}
        \label{subfig:MM1B_mt_rec_error}
    \end{subfigure}
    \caption{When $\lambda = 2$ and $\mu=1$.}
    \label{fig:MM1B_mt_comparison}
\end{figure}

\begin{figure}
    \centering
    \begin{subfigure}[b]{0.24\textwidth}
        \centering
        \includegraphics[width=\linewidth]{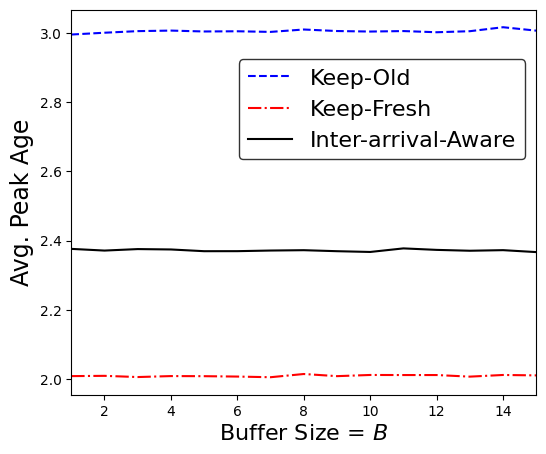}
        \caption{Average peak age.}
        \label{subfig:MM1B_ht_peak_age}
    \end{subfigure}
    \hfill
    \begin{subfigure}[b]{0.24\textwidth}
        \centering
        \includegraphics[width=\linewidth]{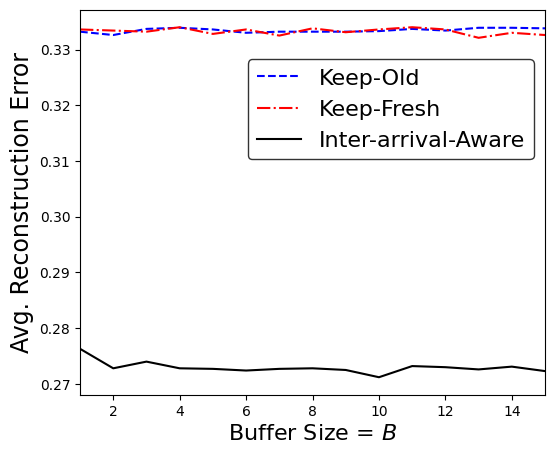}
        \caption{Average reconstruction error.}
        \label{subfig:MM1B_ht_rec_error}
    \end{subfigure}
    \caption{When $\lambda = 200$ and $\mu=1$.}
    \label{fig:MM1B_ht_comparison}
\end{figure}

In contrast, when traffic increases, the impact of the dropping policy becomes more pronounced. With $\lambda=2$ (Fig.~\ref{fig:MM1B_mt_comparison}), buffer overflow occurs more frequently, and the choice of which packets to drop leads to noticeable performance differences between the dropping policies. Under even higher traffic with $\lambda=200$ (Fig.~\ref{fig:MM1B_ht_comparison}), buffer occupancy changes rapidly, and the choice of retained packets has a substantial impact. These results collectively demonstrate that the optimal packet-dropping policy depends on the system objective. Keep-Fresh is preferred when minimizing information latency, while Inter-arrival-Aware is beneficial when preserving historical state information for reconstruction.

\section{Conclusion} \label{sec:conclusion}
In this paper, we investigated the trade-off between real-time monitoring and historical trajectory reconstruction in remote tracking systems. While age-optimal policies minimize AoI by prioritizing freshness, they often discard older packets critical for accurate trajectory reconstruction. To address this, we used reconstruction error as a complementary metric to AoI and analyzed the impact of different packet-dropping policies. Our results showed that the Keep-Fresh policy achieves the lowest AoI but does not necessarily optimize reconstruction accuracy. To balance these objectives, we proposed the Inter-arrival-Aware dropping policy, which dynamically selects packets based on their generation times. Numerical evaluations demonstrated that this policy improves reconstruction accuracy while maintaining reasonable freshness. These findings provide a framework for efficient information management in remote tracking systems, particularly in IoT applications requiring both real-time updates and historical analysis. Future work could explore multi-source networks and learning-based adaptive policies for further optimization.

\bibliographystyle{IEEEtran}
\bibliography{sunjung_remote_tracking}

\appendices
\section{Analytical Evaluation of Reconstruction Error for the Wiener Process} \label{appendix:Wiener}

Let $W(\cdot)=\{W(t)\}_{t\ge 0}$ be a standard Wiener process with mean zero and covariance $\mathbb{E}[W(s)W(u)]=\min\{s,u\}$. Fix two generation times $t_i<t_{i+1}$ and any intermediate time $t\in(t_i,t_{i+1})$. The three–dimensional random vector $(W(t_i),W(t),W(t_{i+1})$ is jointly Gaussian. Hence, the conditional expectation 
\begin{equation}
    \hat{W}(t)=\mathbb{E}[W(t)\mid W(t_i),W(t_{i+1})]
\end{equation}
is the optimal estimator in the sense that it minimizes the mean squared error among all possible estimators. Moreover, since the Wiener process is Gaussian, this conditional expectation is also a linear function of the observations, i.e., it is equivalent to the linear minimum mean square error (LMMSE) estimator.

Let $X=W(t)$ and $Y=(W(t_i),W(t_{i+1}))^{\top}$. For Gaussian vectors, the LMMSE estimator is obtained via linear regression:
\begin{equation}
    \hat{W}(t)=\boldsymbol{\Sigma}_{XY}\boldsymbol{\Sigma}_{YY}^{-1}Y, \label{eq:w_hat}
\end{equation}
where $\boldsymbol{\Sigma}_{YY}=\mathbb{E}[YY^{\top}]$ and $\boldsymbol{\Sigma}_{XY}=\mathbb{E}[XY^{\top}]$. Using the covariance of the Wiener process, we have
\begin{equation}
    \boldsymbol{\Sigma}_{YY}= \begin{pmatrix}
        t_i & t_i\\
        t_i & t_{i+1}
        \end{pmatrix},
    \qquad
    \boldsymbol{\Sigma}_{XY}=\bigl(t_i,\;t\bigr).
\end{equation}
Since the determinant of $\boldsymbol{\Sigma}_{YY}$ is $t_i(t_{i+1}-t_i)$, we have
\begin{equation}
    \boldsymbol{\Sigma}_{YY}^{-1}=
    \frac{1}{t_i(t_{i+1}-t_i)}
        \begin{pmatrix}
            t_{i+1} & -t_i\\
            -t_i & t_i
        \end{pmatrix},
\end{equation}
and thus
\begin{equation}
    \boldsymbol{\Sigma}_{XY}\,\boldsymbol{\Sigma}_{YY}^{-1}
=\frac{1}{t_{i+1}-t_i}\bigl(t_{i+1}-t,\;t-t_i\bigr).
\end{equation}

Substituting into (\ref{eq:w_hat}), we have
\begin{align}
    \hat{W}(t)&=\frac{t_{i+1}-t}{t_{i+1}-t_i}\,W(t_i)
           +\frac{t-t_i}{t_{i+1}-t_i}\,W(t_{i+1}) \nonumber \\
           &=W(t_i)
            +\frac{t-t_i}{t_{i+1}-t_i}\bigl(W(t_{i+1})-W(t_i)\bigr),
\end{align}
for $t_i<t<t_{i+1}$.

The linear–interpolation form of $\hat{W}(t)$ completely specifies the estimator on every inter‑delivery interval $[t_i,t_{i+1}]$. To quantify its accuracy we next evaluate the reconstruction error (RE)
\begin{equation}
    \text{RE}(T)=\frac{1}{\tilde{T}}\int_{0}^{\tilde{T}}\!\bigl(W(s)-\hat{W}(s)\bigr)^{2}ds, 
\end{equation}
where $\tilde{T}=\max\{t_k':t_k'\le T\}$, and $t_k'$ are the packet‑delivery times and $\tilde{T}$ is the last delivery time not exceeding $T$. Because $\hat{W}(t)$ interpolates linearly between $W(t_i)$ and $W(t_{i+1})$, the estimation error $E(t)=W(t)-\hat{W}(t)$ on $[t_i,t_{i+1}]$ is a Brownian bridge that starts and ends at zero. A Brownian bridge of length $L_i=t_{i+1}-t_i$ has conditional variance 
\begin{equation}
    \text{Var}(E(t)| W(t_i),W(t_{i+1}))=\frac{(t-t_i)(t_{i+1}-t)}{L_i}.
\end{equation}
Integrating this variance over the interval and noting that the result is independent of the endpoint values gives
\begin{equation}
    \int_{t_i}^{t_{i+1}}\mathbb{E}[E^{2}(s)]~ds =\frac{L_i^{2}}{6}.
\end{equation}

The intervals between successive deliveries are disjoint, so the total integrated squared error accumulated up to $\tilde{T}$ is the sum of the individual contributions.  Letting $n(\tilde{T})$ for the number of packets delivered by $\tilde{T}$ and setting $t_0=0$, we obtain
\begin{equation}
    \int_{0}^{\tilde{T}}(W(s)-\hat{W}(s))^{2}~ds =\sum_{k=0}^{n(\tilde{T})-1}\frac{(t_{k+1}-t_k)^{2}}{6}.
\end{equation}
Dividing by $\tilde{T}$ yields the analytical expression for the reconstruction error:
\begin{equation}
    \text{RE}(T)=\frac{1}{\tilde{T}}
             \sum_{k=1}^{n(\tilde{T})}\frac{(t_{k}-t_{k-1})^{2}}{6}.
\end{equation}

\section{Proof of Lemma~\ref{lemma:Keep_old_rec_error}} \label{appendix:Keep_old_rec_error}

% \begin{figure}
%     \centering
%     \includegraphics[width=\linewidth]{Figures/Fig_no_replacement_time_line.png}
%     \caption{Sample path of packet arrivals and departures under the Keep-Old policy.}
%     \label{fig:MM12_keep_old_time_line}
% \end{figure}

As discussed in Section~\ref{subsec:RE}, the average reconstruction error $\overline{\text{RE}}$ depends on the second moment $\mathbf{E}[Z_i^2]$ of the inter-arrival times between successfully delivered packets. Let $\tau_{i}$ denote the service start time of packet $i$, and let $W_{i}$ denote the waiting time of packet $i$ in the buffer, which is then given by $W_{i} = \tau_{i} - t_{i}$. Further, let $X_i^*$ denote the time interval between the service start time of $\tau_{i-1}$ and the arrival $t_{i}$ of packet $i$, i.e., $X_i^* = t_{i}-\tau_{i-1}$. Then, as shown in Fig.~\ref{fig:MM12_keep_old_time_line}, the inter-arrival time $Z_i$ is given by $Z_i = W_{i-1} + X^*_i$. Since $W_{i-1}$ and $X^*_i$ are independent, the second moment is given by
\begin{equation}
     \mathbb{E}[Z_i^2] = \mathbb{E}[W_{i-1}^2] + 2\mathbb{E}[W_{i-1}]\mathbb{E}[X^*_i] + \mathbb{E}[(X^*_i)^2].
\end{equation}

Let $\psi_n$ denote the event that when a packet arrives, the system has $n$ packets. From the PASTA (Poisson Arrival See Time Average) property, $\mathbb{P}(\psi_n) = \pi_n$. In an $M/M/1/2$ queue, the transmission probability is given by $\mathbb{P}(\text{tx}) = \pi_0 + \pi_1$. Further, under the Keep-Old policy, packets arriving under the event $\psi_0$ and $\psi_1$ are guaranteed to be served. Thus, we have 
\begin{equation}
    \mathbb{P}(\psi_0 | \text{tx}) = \frac{\mu}{\lambda+\mu} \text{ and } \mathbb{P}(\psi_1 | \text{tx}) = \frac{\lambda}{\lambda+\mu}. \label{eq:app_MM12_psi}
\end{equation}

Under the event $\psi_0$, the arriving packet is immediately served, and thus waiting time is zero. Under the event $\psi_1$, the arriving packet must waiting the amount time of the remaining service time of the in-service packet. Due to the memoryless property of service times, it follows an exponential distribution with rate $\mu$. Let $W$ denote the waiting time for an arbitrary packet. Then, we have $\mathbb{E}[W |\psi_1,\text{tx}] = \frac{1}{\mu}$ and $\mathbb{E}[W^2|\psi_1,\text{tx}] = \frac{2}{\mu^2}$. Combining with (\ref{eq:app_MM12_psi}), we have
\begin{equation}
    \mathbb{E}[W|\text{tx}] = \frac{\lambda}{\mu(\lambda+\mu)} \text{ and } \mathbb{E}[W^2|\text{tx}] = \frac{2\lambda}{\mu^2(\lambda+\mu)}.
\end{equation}

Further, due to the memoryless property of inter-arrival times, the time interval $X^*_i$ follows an exponential distribution with rate $\lambda$, and thus we have $\mathbf{E}[X^*_i]=\frac{1}{\lambda}$ and $\mathbb{E}[(X^*_i)^2] = \frac{2}{\lambda^2}$. Hence, the second moment is given by 
\begin{equation}
    \begin{split}
        \mathbb{E}[Z_i^2] &= \frac{2\lambda}{\mu^2(\lambda+\mu)} + \frac{2}{\mu(\lambda+\mu)} + \frac{2}{\lambda^2} = \frac{2}{\mu^2} + \frac{2}{\lambda^2}.
    \end{split}   
\end{equation}

Combining with (\ref{eq:effective_arrival_rate}), we can obtain
\begin{equation}
    \overline{RE}_{\text{Keep-Old}} =\frac{(\lambda+\mu)(\lambda^2+\mu^2)}{3\lambda\mu(\lambda^2+\lambda\mu+\mu^2)}.
\end{equation}

%---------------------------------
\section{Proof of Lemma~\ref{lemma:Keep_fresh_rec_error}} \label{appendix:Keep_fresh_rec_error} 
% \begin{figure}
%     \centering
%     \begin{subfigure}[t]{\linewidth}
%         \centering
%         \includegraphics[width=\linewidth]{Figures/Fig_replacement_time_line_1.png}
%         \caption{No arrival occurs during the service time of packet $i_{k-1}$.}
%         \label{subfig:replacment_time_line1}
%     \end{subfigure}
%     \hfill
%     \begin{subfigure}[t]{\linewidth}
%         \centering     
%         \includegraphics[width=\linewidth]{Figures/Fig_replacement_time_line.png}
%         \caption{Arrivals occur during the service time of packet $i_{k-1}$.}
%         \label{subfig:replacment_time_line2}
%     \end{subfigure}
%     \caption{Sample path of packet arrivals and departures under the Keep-Fresh policy.}
%     \label{fig:replacment_time_line}
% \end{figure}

In this section, we analyze the average reconstruction error under the Keep-Fresh policy. Let $I_i$ denote the time interval between the service start time $\tau_{i-1}$ of packet $i$ and the arrival time $t_{i}$ of packet $i$ as shown in Fig.~\ref{fig:replacment_time_line}. Then, the inter-arrival time $Z_i$ can be expressed as $Z_i = W_{i-1}+I_{i}$, and its second moment can be expressed as
\begin{equation}
    \mathbb{E}[Z_i^2] = \mathbb{E}[W_{i-1}^2] + 2\mathbb{E}[W_{i-1}]\mathbb{E}[I_i] + \mathbb{E}[I_i^2], \label{eq:app_EZ2}
\end{equation}
where $W_{i-1}$ and $I_i$ are independent due to the memoryless property of inter-arrival and service times.

\subsubsection{The Waiting Time $W_{i-1}$}

Unlike the Keep-Old policy, where a packet arriving at $\psi_1$ is guaranteed to be served and at $\psi_2$ is always dropped, the Keep-Fresh policy allows packets arriving at either $\psi_1$ or $\psi_2$ to first be stored in the buffer. These packets can then either be served or dropped depending on whether new arrivals occur during the remaining service time of the in-service packet. Let $\phi_r$ denote the event that no arrival occurs during the remaining service time $R$ of the in-service packet. This event is the equivalent to the event $(\text{tx} | \psi_1 \text{ or } \psi_2)$. The probability $\mathbb{P}(\phi_r)$ of this event is given by
\begin{align}
    \mathbb{P}(\phi_r) &= \int_0^\infty \mathbb{P}(\phi|R=r)f_R(r)dr  \nonumber \\
    &= \int_0^\infty \frac{(\lambda r)^0 e^{-\lambda r}}{0!} \mu e^{-\mu r}  dr = \frac{\mu}{\lambda+\mu}, \label{eq:app_p_no_arrival}
\end{align}
where $f_R(r)$ is the probability density function of the remaining service time $R$.

The waiting time $W$ for an arbitrary packet, conditioned on $(\psi_1 \text{ or } \psi_2, \text{tx})$, is equivalent to the remaining service time $R$ conditioned on $\phi_r$. Then, we have
\begin{align}
    f_W(w|\psi_1 \text{ or } \psi_2, \text{tx}) &= f_R(w|\phi_r) = \frac{\mathbb{P}(\phi_r|R=w)f_R(w)}{\mathbb{P}(\phi_r)} \nonumber \\
    &= (\lambda+\mu)e^{-(\lambda+\mu)w}. \label{eq:pdf_W}
\end{align}
The probability that the packet arrives when the server is busy and is successfully transmitted is
\begin{align}
    \mathbb{P}(\psi_1 \text{ or } \psi_2,\text{tx}) &= \mathbb{P}(\psi_1 \text{ or } \psi_2)\mathbb{P}(\text{tx} | \psi_1 \text{ or } \psi_2) \nonumber \\ 
    &= (1-\pi_0)\frac{\mu}{\lambda+\mu}.
\end{align}
Since a packet arriving at $\psi_0$ is always served, we have $\mathbb{P}(\text{tx}) = \pi_0 + \mathbb{P}(\psi_1 \text{ or } \psi_2,\text{tx})$, and thus $\mathbb{P}(\psi_1 \text{ or } \psi_2|\text{tx}) = \mathbb{P}(\psi_1 \text{ or } \psi_2,\text{tx})/\mathbb{P}(\text{tx}) = \frac{\lambda}{\lambda+\mu}$. Hence, we can obtain 
\begin{equation}
    \mathbb{E}[W_{i-1}] = \frac{\lambda}{(\lambda+\mu)^2} \text{ and } \mathbb{E}[W_{i-1}^2] = \frac{2\lambda}{(\lambda+\mu)^3}. \label{eq:app_EWEW2}
\end{equation}

\subsubsection{The Time Interval $I_i$}
Consider the event $\phi_s$, where no new arrival occurs during the service time $S_{i-1}$ of packet $i-1$ as shown in Fig.~\ref{fig:replacment_time_line}(\subref{subfig:replacment_time_line1}). Under this condition, packet $i$ becomes the first packet to arrive after the departure of packet $i-1$. In this case, the time interval $I_i$ can be expressed as $I_i = S_{i-1}+X_i^*$, where $S_{i-1}$ is the service time of packet $i-1$ and $X_i^* = t_{i}-t'_{i-1}$ is the time interval between the departure of packet $i-1$ and the arrival of packet $i$. 

Due to the memoryless property of inter-arrival times, $X_i^*$, conditioned on $\phi_s$, follows an exponential distribution with rate $\lambda$. Thus, we have $\mathbb{E}[X_i^*|\phi_s] = \frac{1}{\lambda}$ and $\mathbb{E}[(X_i^*)^2|\phi_s] = \frac{2}{\lambda^2}$. Further, following the same line of reasoning used to derive $f_R(r|\phi_s)$ in (\ref{eq:pdf_W}), the conditional pdf of $S_{i-1}$ given $\phi_s$ is given by
\begin{align}
    f_{S_{i-1}}(s|\phi_s) &= (\lambda+\mu)e^{-(\lambda+\mu)s}. \label{eq:app_pdf_S}
\end{align}
Thus, we have 
\begin{equation}
    \mathbb{E}[S_{i-1}|\phi_s] = \frac{1}{\lambda+\mu} ~\text{ and }~ \mathbb{E}[S^2_{i-1}|\phi_s] = \frac{2}{(\lambda+\mu)^2}.
\end{equation}
Finally, we have
\begin{equation}
    \mathbb{E}[I_i | \phi_s] =\mathbb{E}[S_{i-1}|\phi_s]+\mathbb{E}[X_i^*|\phi_s]  = \frac{2\lambda+\mu}{\lambda(\lambda+\mu)}, \label{eq:app_EI_0}
\end{equation}
and
\begin{align}
     % \mathbb{E}[I_i^2 | \phi_s] &\stackrel{(A)}{=}  \mathbb{E}[S^2_{i-1}|\phi_s] + 2\mathbb{E}[S_{i-1}|\phi_s]\mathbb{E}[X_i^*|\phi_s]+\mathbb{E}[(X_i^*)^2|\phi_s] \nonumber \\
     \mathbb{E}[I_i^2 | \phi_s] &\stackrel{(A)}{=}  \mathbb{E}[S^2_{i-1}|\cdot] + 2\mathbb{E}[S_{i-1}|\cdot]\mathbb{E}[X_i^*|\cdot]+\mathbb{E}[(X_i^*)^2|\cdot] \nonumber \\
     % &=\frac{2}{(\lambda+\mu)^2} + \frac{2}{\lambda(\lambda+\mu)} + \frac{2}{\lambda^2} 
     &= \frac{2(3\lambda^2+3\lambda\mu+\mu^2)}{\lambda^2(\lambda+\mu)^2}, \label{eq:app_EI2_0}
\end{align}
where $(A)$ comes from the independence of $S_{i-1}$ and $X_i^*$ due to the memoryless property of service times and inter-arrival times. The condition $\phi_s$ is omitted for brevity in the notation.

Now, consider the event $\bar{\phi}_s$, where new arrivals occur during the service time $S_{i-1}$ of packet $i-1$ as shown in Fig.~\ref{fig:replacment_time_line}(\subref{subfig:replacment_time_line2}). Under this condition, only the latest arriving packet will be served, while any other arrivals will be dropped. Let $\bar{\phi}_{s,m}$ denote the sub-event under $\bar{\phi}_s$ where exactly $m$ arrivals occur during the service time $S_{i-1}$. 

Let $X_1$ be the time interval between the service start time $\tau_{i-1}$ of packet $i-1$ and the first packet arrival. For $j=2,...,m$, let $X_j$ be the inter-arrival time between the $(j-1)^{th}$ and $j^{th}$ arrival, where $m^{th}$ arrival is packet $i$. We then define the cumulative sum of these inter-arrival times as 
\begin{equation}
    \Sigma_{m+1} = \sum_{j=1}^{m+1} X_{j}.
\end{equation}
This sum follows an Erlang distribution, of which pdf is given by
\begin{equation}
    f_{\Sigma_{m}}(s) = \frac{\lambda^{m}s^{m-1}e^{-\lambda s}}{(m-1)!}, ~s\ge 0, \label{eq:app_f_pdf}
\end{equation}
and its cdf is given by
\begin{equation}
    F_{\Sigma_{m}}(s) = 1-e^{-\lambda s} \sum_{k=0}^{m-1} \frac{(\lambda s)^k}{k!}, ~s\ge 0.
\end{equation}

Let $\phi_r$ denote the event that no arrival occurs during the remaining service time after the arrival time $t_{i}$ of packet $i$. The probability of $\bar{\phi}_{s,m}$ is given by $\mathbb{P}(\bar{\phi}_{s,m}) = \mathbb{P}(\Sigma_{m} < S_{i-1}, \phi_r) = \mathbb{P}(\Sigma_{m} < S_{i-1})\mathbb{P}(\phi_r)$ due to the memoryless property of service time. From (\ref{eq:app_p_no_arrival}), we have $\mathbb{P}(\phi_r) = \frac{\mu}{\lambda+\mu}$. 
Further, we have
\begin{align}
        \mathbb{P}(\Sigma_{m} < S_{i-1}) &= \int_0^\infty \mathbb{P}(\Sigma_{m} < t | S_{i-1}=t) f_{S_{i-1}}(t) ~dt \nonumber\\
        &\hspace{-1cm}= \int_0^\infty \left(1-e^{-\lambda t} \sum_{k=0}^{m-1} \frac{(\lambda t)^k}{k!}\right)\mu e^{-\mu t} ~dt \nonumber\\
        % &= 1- \int_0^\infty \sum_{k=0}^{m-1} \frac{(\lambda t)^k}{k!} \mu e^{-(\lambda +\mu)t} ~dt \nonumber\\
        &\hspace{-1cm}\stackrel{(A)}{=} 1-  \sum_{k=0}^{m-1}\int_0^\infty \frac{(\lambda t)^k}{k!} \mu e^{-(\lambda +\mu)t} ~dt \nonumber\\
        &\hspace{-1cm}\stackrel{(B)}{=} 1- \frac{\mu}{\lambda+\mu} \sum_{k=0}^{m-1} \left(\frac{\lambda}{\lambda+\mu}\right)^k \nonumber\\
        &\hspace{-1cm}= 1-\left(1-\left(\frac{\lambda}{\lambda+\mu}\right)^{m}\right)  = \left(\frac{\lambda}{\lambda+\mu}\right)^{m}, \label{eq:app_prob_ST_S}
\end{align}
where $(A)$ comes from the Fubini's theorem, and $(B)$ from the Laplace transform:
\begin{equation}
    \int_0^\infty t^k e^{-(\lambda+\mu)t} ~dt = \frac{k!}{(\lambda+\mu)^{k+1}}.
\end{equation}
Hence, we have
\begin{equation}
    \mathbb{P}(\bar{\phi}_{s,m}) = \left(\frac{\lambda}{\lambda+\mu}\right)^{m} \frac{\mu}{\lambda+\mu}. \label{eq:app_psi_n_prob}
\end{equation}

We now obtain $\mathbb{E}[\Sigma_{m} | \bar{\phi}_{s,m}]$ as
\begin{align}
        \mathbb{E}[\Sigma_{m} | \bar{\phi}_{s,m}] &=\mathbb{E}[\Sigma_{m} | \Sigma_{m} < S_{i-1},\phi] \nonumber\\ 
        &= \frac{\int_0^\infty s \mathbb{P}(\Sigma_{m} < S_{i-1} | \Sigma_{m}=s) f_{\Sigma_{m}}(s) ~dt}{\mathbb{P}(\Sigma_{m} < S_{i-1})}, \nonumber \\
        &\stackrel{(A)}{=} \frac{\int_0^\infty s e^{-\mu s} \frac{\lambda^{m}s^{n}e^{-\lambda s}}{(m-1)!}  ~dt}{\mathbb{P}(\Sigma_{m} < S_{i-1})}  \stackrel{(B)}{=} \frac{\frac{\lambda^{m}m!}{(\lambda+\mu)^{m+1}}}{\left(\frac{\lambda}{\lambda+\mu}\right)^{m}} \nonumber \\
        &= \frac{m}{\lambda+\mu}, \label{eq:app_EST_phi_m}
\end{align}
where $(A)$ comes from (\ref{eq:app_f_pdf}) and the fact that $\mathbb{P}(\Sigma_{m} < S_{i-1} | \Sigma_{m}=s)=\mathbb{P}(S_{i-1}>s) = e^{-\mu s}$, and $(B)$ from (\ref{eq:app_prob_ST_S}) and the Laplace transform:
\begin{equation}
    \int_0^\infty s^{m} e^{-(\lambda+\mu)s} ~ds = \frac{m!}{(\lambda+\mu)^{m+1}}.
\end{equation}
Similarly, we can obtain 
\begin{align}
    \mathbb{E}[\Sigma^2_{m} | \bar{\phi}_{s,m}] &= \mathbb{E}[\Sigma^2_{m} | \Sigma_{m} < S_{i-1}] \nonumber\\
    &= \frac{\int_0^\infty s^2 \mathbb{P}(\Sigma_{m} < S_{i-1} | \Sigma_{m}=s) f_{\Sigma_{m}}(s) ~dt}{\mathbb{P}(\Sigma_{m} < S_{i-1})} \nonumber \\
    &= \frac{m(m+1)}{(\lambda+\mu)^2}. \label{eq:app_EST2_eta_n}
\end{align}
Hence, from (\ref{eq:app_EI_0}), (\ref{eq:app_psi_n_prob}) and (\ref{eq:app_EST_phi_m}), we have
\begin{align}
    \mathbb{E}[I_i]&= \mathbb{P}(\phi)\mathbb{E}[I_i|\phi]+\sum_{m=1}^\infty \mathbb{P}(\bar{\phi}_{s,m})\mathbb{E}[I_i|\bar{\phi}_{s,m}] \nonumber\\
    &= \frac{\mu}{\lambda+\mu}\frac{2\lambda+\mu}{\lambda(\lambda+\mu)} + \sum_{m=1}^\infty \left(\frac{\lambda}{\lambda+\mu}\right)^{m} \frac{\mu}{\lambda+\mu} \frac{m}{\lambda+\mu} \nonumber \\
    &=  \frac{\mu(2\lambda+\mu)}{\lambda(\lambda+\mu)^2} + \frac{\lambda}{\mu(\lambda+\mu)}. \label{eq:app_EI_total}
\end{align}
Similarly, from (\ref{eq:app_EI2_0}), (\ref{eq:app_psi_n_prob}) and (\ref{eq:app_EST2_eta_n}), we have
\begin{align} 
\mathbb{E}[I^2_i]&= \mathbb{P}(\phi)\mathbb{E}[I^2_i|\phi]+\sum_{m=1}^\infty \mathbb{P}(\bar{\phi}_{s,m})\mathbb{E}[I^2_i|\bar{\phi}_{s,m}] \nonumber\\
    &= \frac{\mu}{\lambda+\mu} \frac{2(3\lambda^2+3\lambda\mu+\mu^2)}{\lambda^2(\lambda+\mu)^2} \nonumber \\
    &\hspace{1.8cm}+ \sum_{m=1}^\infty \left(\frac{\lambda}{\lambda+\mu}\right)^{m} \frac{\mu}{\lambda+\mu} \frac{m(m+1)}{(\lambda+\mu)^2} \nonumber \\
    &= \frac{2\mu(3\lambda^2+3\lambda\mu+\mu^2)}{\lambda^2(\lambda+\mu)^3}+\frac{2\lambda}{\mu^2(\lambda+\mu)}. \label{eq:app_EI2_total}
\end{align}

Combining (\ref{eq:app_EZ2}), (\ref{eq:app_EWEW2}), (\ref{eq:app_EI_total}) and (\ref{eq:app_EI2_total}), we have
\begin{align}
    \mathbb{E}[Z_i^2] &= \frac{2\lambda}{(\lambda+\mu)^3} + \frac{2\lambda}{(\lambda+\mu)^2} \left( \frac{\lambda}{\mu(\lambda+\mu)} + \frac{\mu(2\lambda+\mu)}{\lambda(\lambda+\mu)^2}\right) \nonumber \\
    &\hspace{0.5cm}+ \frac{2\lambda}{\mu^2(\lambda+\mu)} + \frac{2\mu(3\lambda^2+3\lambda\mu+\mu^2)}{\lambda^2(\lambda+\mu)^3} 
    % &= \frac{2\lambda\mu}{(\lambda+\mu)^4} + \frac{2(\lambda^2+\mu^2)}{\lambda^2\mu^2}-\frac{4\lambda}{(\lambda+\mu)^3}. \label{eq:app_EZ2}
\end{align}
Rearranging this, we can obtain the long-term average reconstruction error reconstruction error $\overline{RE}_{\text{Keep-Fresh}}$ under the Keep-Fresh policy, which is given by 
\begin{equation}
    \overline{RE}_{\text{Keep-Fresh}} = \frac{\lambda_{\text{eff}}\mathbb{E}[Z_i^2]}{6},
\end{equation}
where
\begin{equation}
    \lambda_{\text{eff}} = \frac{\lambda(\mu^2+\lambda\mu)}{\mu^2 + \lambda\mu+\lambda^2},
\end{equation}
and
\begin{equation}
    \mathbb{E}[Z_i^2] = \frac{2\lambda\mu}{(\lambda+\mu)^4} + \frac{2(\lambda^2+\mu^2)}{\lambda^2\mu^2}-\frac{4\lambda}{(\lambda+\mu)^3}. 
\end{equation}

\section{Proof of Lemma~\ref{lemma:IaA_age}} \label{appendix:IaA_age}

We first consider a single service interval $[\tau_i,\tau_i+S]$, where $S\sim \text{exp}(\mu)$ is the service duration. After normalizing time so that this interval becomes $[0,1]$, each new arrival lies (conditionally) in $(0,1)$. By the order‐statistics property of Poisson arrivals, when $k$ arrivals occur within a service period of length $S$, their epochs are distributed as the order statistics of $k$ i.i.d. $\text{Uniform}[0,S]$ random variables; dividing by $S$ renders them uniform in $(0,1)$. 
In the heavy-traffic regime, given the waiting time $W_{i-1}$ of the in-service packet, the waiting time $W_{i}$ of the buffered packet approaches to $1-2^{\alpha}W_{i-1}$, where
\begin{equation}
    \alpha = \argmin_{\alpha=0,1,2,...} \{2^\alpha W_{i-1} > 0.5 \}.
\end{equation}

We show that this recurrence has a unique limiting distribution in $(0,1)$. Let us define the map 
\begin{equation}
    F ~:~ (x,y) \mapsto \left(y,1-2^{\alpha(x)}y\right),
\end{equation}
where $\alpha(x) = \min\{\alpha\ge 0 : 2^\alpha x > 0.5\}$ with the domain $\mathcal{D}=(0,1)\times (0,1)$. Then, we have $(W_{i-1},W_i) \mapsto (W_i, W_{i+1}) = F(W_{i-1},W_i)$.  For each integer $\alpha \ge 0$, we define
\begin{equation}
    I_\alpha = \left(\frac{0.5}{2^{\alpha}},\frac{0.5}{2^{\alpha-1}}\right],
\end{equation}
on which of each $\alpha(x) = \alpha$. Hence $F$ is piecewise linear, with each piece sending $(x,y)\mapsto(y,1-\alpha y)$.

We first show the existence of an invariant measure. Since $F$ is defined on the compact set $\bar{D}\in [0,1]^2$ and is piecewise continuous (affine on each subdomain corresponding to a fixed $\alpha$), a standard Markov chain compactness argument guarantees the existence of at least one invariant Borel probability measure~\cite{meyn2012markov}. Concretely, for any initial 
$(x,y)\in\mathcal{D}$, we form the empirical distribution of the orbit $(F^k(x,y))_{k \ge 1}$, where the set of such empirical measures is sequentially compact, and any limit of those measures is invariant under $F$. Therefore at least one invariant measure $\mu^*$ exists.

We then show the uniqueness and ergodicity. The map $F$ is piecewise affine and has the property that almost every orbit under $F$ is dense in $\mathcal{D}$. In particular, $F$ is topologically transitive and the invariant measure is unique~\cite{boyarsky2012laws}. Hence, any invariant measure must be unique, and the marginal distribution of $\{W_i\}$ in steady state is uniquely determined.

Since $\alpha(x)$ defines infinitely many subintervals in the $x$-direction, obtaining a closed-form expression for the invariant measure is intractable.  Instead, one can approximate the unique invariant measure numerically by discretizing the state space and iterating the two-dimensional map
\begin{equation}
    (W_{i-1},W_i)\mapsto (W_i,\,1-2^{\alpha(W_{i-1})}W_i).
\end{equation}
Numerical experiments using this procedure yield that the steady-state distribution of $W$ has a mean of approximately $0.375$, which in turn implies that the average waiting time is numerically close to $0.375S$, where $S$ is the service duration.

We recall that, from $(\ref{eq:peak_age_expression})$ the average peak age can be expressed as $\Bar{A} = \mathbb{E}[A_k] = \mathbb{E}[T_{i_{k-1}}] + \mathbb{E}[Y_k]$, where $T_{i_{k-1}}$ is the system time of packet $i_{k-1}$ and $Y_k$ is the inter-departure time between packets $i_{k-1}$ and $i_k$. The expected system time can be written as $\mathbb{E}[T_{k-1}] = \mathbb{E}[T_k] = \mathbb{E}[W_{i_k}]+\mathbb{E}[S_{i_k}]$, with $\mathbb{E}[S_{i_k}] = \frac{1}{\mu}$ since the service time follows an exponential distribution with rate $\mu$. In the heavy traffic regime, the buffer is almost always non-empty, thus we have $\mathbb{E}[Y_k]  \rightarrow \frac{1}{\mu}$ as $\lambda\rightarrow\infty$. Hence, the average peak age converges to $\bar{A}\rightarrow\frac{1}{\mu} +\frac{1}{\mu} + \frac{0.375}{\mu} = \frac{2.375}{\mu}$.

\section{Proof of Lemma~\ref{lemma:MM1B_keep_old_age}} \label{appendix:MM1B_keep_old_age}
In other to show Lemma IV.1, we recall that, 
\begin{equation}
    \Bar{A} = \mathbb{E}[T_{i_{k-1}}] + \mathbb{E}[Y_k]. \label{eq:peak_age_expression}
\end{equation}
%from $(\ref{eq:peak_age_expression})$ the average peak age can be expressed as $\Bar{A} = \mathbb{E}[A_k] = \mathbb{E}[T_{i_{k-1}}] + \mathbb{E}[Y_k]$, 
where $T_{i_{k-1}}$ is the system time of packet $i_{k-1}$ and $Y_k$ is the inter-departure time between packets $i_{k-1}$ and $i_k$. The expected system time can be written as $\mathbb{E}[T_{k-1}] = \mathbb{E}[T_k] = \mathbb{E}[W_{i_k}]+\mathbb{E}[S_{i_k}]$, with $\mathbb{E}[S_{i_k}] = \frac{1}{\mu}$ since the service time follows an exponential distribution with rate $\mu$.

Further, we recall the results obtained the proofs of Lemma III.1 and III.2. Let $\psi_n$ denote the event that when a packet arrives, the system has $n$ packets. From the PASTA (Poisson Arrival See Time Average) property, $\mathbb{P}(\psi_n) = \pi_n$.  Let $\phi_r$ denote the event that no arrival occurs during the remaining service time $R$ of the in-service packet. This event is the equivalent to the event $(\text{tx} | \psi_1 \text{ or } \psi_2)$. The probability $\mathbb{P}(\phi_r)$ of this event is given by
\begin{align}
    \mathbb{P}(\phi_r) &= \int_0^\infty \mathbb{P}(\phi|R=r)f_R(r)dr  \nonumber \\
    &= \int_0^\infty \frac{(\lambda r)^0 e^{-\lambda r}}{0!} \mu e^{-\mu r}  dr = \frac{\mu}{\lambda+\mu}, \label{eq:app_p_no_arrival}
\end{align}
where $f_R(r)$ is the probability density function of the remaining service time $R$.

The waiting time $W$ for an arbitrary packet, conditioned on $(\psi_1 \text{ or } \psi_2, \text{tx})$, is equivalent to the remaining service time $R$ conditioned on $\phi_r$. Then, we have
\begin{align}
    f_W(w|\psi_1 \text{ or } \psi_2, \text{tx}) &= f_R(w|\phi_r) = \frac{\mathbb{P}(\phi_r|R=w)f_R(w)}{\mathbb{P}(\phi_r)} \nonumber \\
    &= (\lambda+\mu)e^{-(\lambda+\mu)w}. \label{eq:pdf_W}
\end{align}
Further, following the same line of reasoning used to derive $f_R(r|\phi_s)$ in (\ref{eq:pdf_W}), the conditional pdf of $S_{i-1}$ given $\phi_s$ is given by
\begin{align}
    f_{S_{i-1}}(s|\phi_s) &= (\lambda+\mu)e^{-(\lambda+\mu)s}. \label{eq:app_pdf_S}
\end{align}

% We recall that, from $(\ref{eq:peak_age_expression})$ the average peak age can be expressed as $\Bar{A} = \mathbb{E}[A_k] = \mathbb{E}[T_{i_{k-1}}] + \mathbb{E}[Y_k]$, where $T_{i_{k-1}}$ is the system time of packet $i_{k-1}$ and $Y_k$ is the inter-departure time between packets $i_{k-1}$ and $i_k$. The expected system time can be written as $\mathbb{E}[T_{k-1}] = \mathbb{E}[T_k] = \mathbb{E}[W_{i_k}]+\mathbb{E}[S_{i_k}]$, with $\mathbb{E}[S_{i_k}] = \frac{1}{\mu}$ since the service time follows an exponential distribution with rate $\mu$. 
We now show Lemma IV.1. The expected service time $\mathbb{E}[S_{i_k}]$ is policy-independent and is determined solely by the exponential distribution of the service times, which has a value of $\frac{1}{\mu}$. To compute $\mathbb{E}[W_{i_k}]$, note that under the event $\psi_0$, the arriving packet is immediately served, giving $\mathbb{E}[W_{i_k}|\psi_0]=0$. Under $\psi_B$, the arriving packet is the freshest and will be served immediately after the in-service packet departs. Thus, the waiting time $W_{i_k}$ conditioned on $\psi_B$ is equivalent to the remaining service time of the in-service packet, which follows an exponential distribution with rate $\mu$. Thus, we have $\mathbb{E}[W_{i_k}|\psi_B] = \frac{1}{\mu}$.

For a packet arriving under $\psi_n$, $n\in\{1,...,B-1\}$, a new arrival during the remaining service time will make it stale under the LCFS Keep-Old policy. Hence, for a packet arriving under the event $\psi_n$, $n\in\{1,...,B-1\}$, to be considered fresh, no arrivals must occur during the remaining service time. From (\ref{eq:app_p_no_arrival}), the probability that a packet under $\psi_n$ is fresh is  $\mathbb{P}(\text{fresh} | \psi_n)=\frac{\mu}{\lambda+\mu}$. Further, from (\ref{eq:pdf_W}), the conditional PDF of the waiting time $W$ given ($\psi_n,\text{fresh}$) is given by
\begin{equation}
    \begin{split}
        f_{W}(w|\psi_n,\text{fresh}) = (\lambda+\mu) e^{-(\lambda+\mu)w}. \label{eq:app_pdf_conditional_w}
    \end{split}
\end{equation}
Hence, we have $\mathbb{E}[W|\psi_n,\text{fresh}] = \frac{1}{\lambda+\mu}$. Combining all the cases, we have
\begin{align}
    \mathbb{E}[W_{i_k}] &= \sum_{n=1}^{B} \mathbb{P}(\psi_n)\mathbb{E}[W_{i_k}|\psi_n] \nonumber \\
    &= \frac{\pi_B}{\mu} + \sum_{n=1}^{B-1} \frac{\pi_n \mu}{\lambda+\mu} \frac{1}{\lambda+\mu}. \label{eq:app_MM1B_keep_old_EW}
\end{align}

\begin{figure}
    \centering
    \begin{subfigure}{0.9\linewidth}
        \centering
        \includegraphics[width=\linewidth]{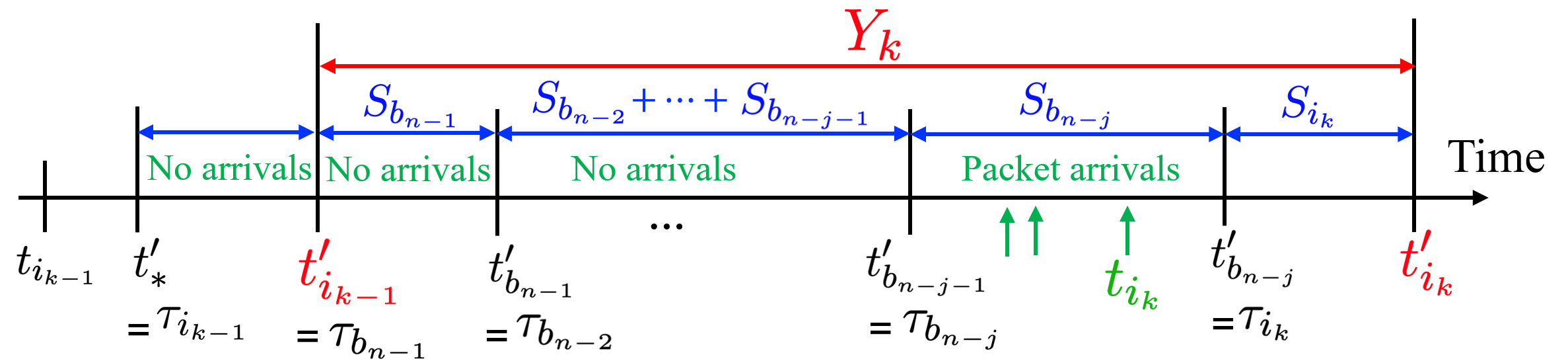}
        \caption{No arrival occurs until the departure of packet $B_{n-j+1}$, and a new arrival occurs during the service time of packet $B_{n-j}$.}
        \label{subfig:MM1B_keep_old_inter_departure_1}
    \end{subfigure}
    \hfill
    \begin{subfigure}{0.9\linewidth}
        \centering     
        \includegraphics[width=\linewidth]{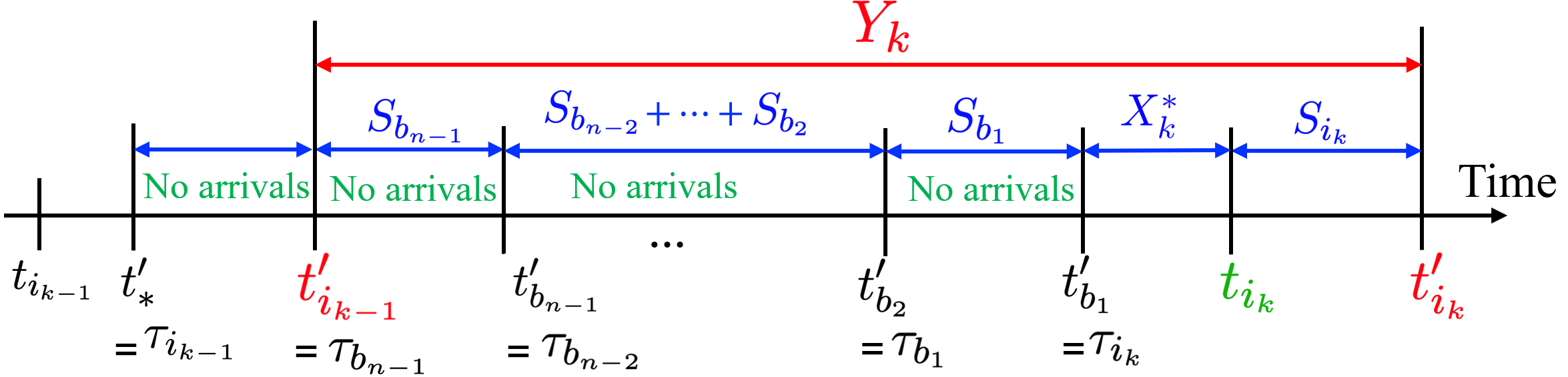}
        \caption{No arrival occurs until the departure of packet $B_1$.}
        \label{subfig:MM1B_keep_old_inter_departure_2}
    \end{subfigure}
    \caption{Sample path of packet arrivals and departures under the Keep-Old policy, where packet $i_k$ arrives when system has $n$ packets (event $\psi_n$).}
    \label{fig:MM1B_keep_old_inter_departure}
\end{figure}

To calculate the expected inter-departure time $\mathbb{E}[Y_k]$, we first consider the case where packet $i_{k-1}$ arrives under $\psi_0$. In this case, packet $i_k$ is fresh, and the next delivered packet will also be fresh due to the LCFS discipline. If new arrivals occur during the service time of packet $i_{k-1}$ with probability $\frac{\lambda}{\lambda+\mu}$, the latest arriving packet $i_{k}$ will be served immediately after the departure of packet $i_{k-1}$. In this case, the inter-departure time between packets $i_{k-1}$ and $i_{k}$ equals the service time of packet $i_{k}$, which has an expected value of $\frac{1}{\mu}$. Otherwise with probability $\frac{\mu}{\lambda+\mu}$, the inter-departure time is the sum of two components: (1) the time interval between the departure of packet $i_{k-1}$ and the arrival of the next packet $i_{k}$, which has an expected value of $\frac{1}{\lambda}$, and (2) the service time of packet $i_{k}$, which has an expected value of $\frac{1}{\mu}$. Thus, the expected inter-departure time conditioned on $\psi_0$ is given by
\begin{equation}
   \textstyle \mathbb{E}[Y_k | \psi_0] = \frac{\lambda}{\lambda+\mu}\frac{1}{\mu} + \frac{\mu}{\lambda+\mu}\left(\frac{1}{\lambda} + \frac{1}{\mu}\right) = \frac{\lambda}{(\lambda+\mu)\mu} + \frac{1}{\lambda}.
\end{equation}

Next, suppose that packet $i_{k-1}$ arrives under the event $\psi_n$, where $n=1,...,B-1$, and is fresh with probability $\frac{\mu}{\lambda+\mu}$ as shown in Fig.~\ref{fig:MM1B_keep_old_inter_departure}, or suppose that packet $i_k$ arrives under the event $\psi_B$, in which case this packet is fresh under the LCFS Keep-Old policy. In both cases, there are $n$ packets in the buffer just before the departure of the in-service packet. Let $b_l$ denote that the $l^{th}$ packet in the buffer, where $l=1$ implies the packet at the head of the buffer. The inter-departure time depends on arrival and service dynamics described by the events $\zeta_0,\zeta_1,...,\zeta_n$:
\begin{itemize}
    \item $\zeta_0$: New arrivals occur during the service time of packet $i_{k-1}$ with probability $\frac{\lambda}{\lambda+\mu}$. In this case, the inter-departure time equals to the service time $S_{i_{k}}$ of the latest arriving packet $i_{k}$, which has an expected value of $\frac{1}{\mu}$.
    \item $\zeta_1$: No arrival occurs during the service time of packet $i_{k-1}$ and new arrivals occur during the service time of packet $b_{n-1}$ with probability $\frac{\mu}{\lambda+\mu}\frac{\lambda}{\lambda+\mu}$. In this case, the inter-departure time equals to the sum of the services time $S_{B_{n-1}}$ and the service time $S_{i_{k}}$, which has an expected value of $\frac{2}{\mu}$.
    \item $\zeta_j$ for $j = 2,...,n-1$: No arrival occurs until the departure of packet $b_{n-j+1}$, and a new arrival occurs during the service time of packet $b_{n-j}$ with probability $\left(\frac{\mu}{\lambda+\mu}\right)^{j}\frac{\lambda}{\lambda+\mu}$. In this case, the inter-departure time equals to the sum of the service times $S_{b_{n-1}}$,...,$S_{b_{n-j}}$ and the service time $S_{i_{k}}$, which has an expected value of $\frac{j+1}{\mu}$ as shown in Fig.~\ref{fig:MM1B_keep_old_inter_departure}(\subref{subfig:MM1B_keep_old_inter_departure_1}).
    \item $\zeta_n$: No arrival occurs until the departure of packet $b_1$ with probability $\left(\frac{\mu}{\lambda+\mu}\right)^{n}$ as shown in Fig.~\ref{fig:MM1B_keep_old_inter_departure}(\subref{subfig:MM1B_keep_old_inter_departure_2}). In this case, the inter-departure time equals to the sum of the service times $S_{b_{n-1}},...,S_{b_1}$, the time interval $X_k^*$ between the departure of packet $B_1$ and the arrival of the next packet $i_{k}$ and the service time $S_{i_{k}}$, which has an expected value of $\frac{n}{\mu}+\frac{1}{\lambda}$.
\end{itemize}
Thus, the expected inter-departure time $Y$ conditioned on $(\psi_n,\text{fresh})$ for $n=1,...,B$ is given by
\begin{align}
    &\mathbb{E}[Y | \psi_n,\text{fresh}] \label{eq:app_MM1B_keep_old_conditional_ET}\\
    &= \sum_{j=0}^{n-1} \left(\frac{\mu}{\lambda+\mu}\right)^{j} \frac{\lambda}{\lambda+\mu} \frac{j+1}{\mu} + \left(\frac{\mu}{\lambda+\mu}\right)^{n} \left(\frac{n}{\mu}+\frac{1}{\lambda}\right). \nonumber
\end{align}
The expected inter-departure time is given by
\begin{equation}
    \begin{split}
        \mathbb{E}[Y_k] &\textstyle = \pi_0 \left(\frac{\lambda}{(\lambda+\mu)\mu} + \frac{1}{\lambda}\right) + \left(\pi_B + \frac{\mu}{\lambda+\mu}\sum_{n=1}^{B-1}\pi_n\right) \\
        &\hspace{-0.4cm}\textstyle\cdot\left(\sum_{j=0}^{n-1} \left(\frac{\mu}{\lambda+\mu}\right)^{j} \frac{\lambda}{\lambda+\mu} \frac{j+1}{\mu} + \left(\frac{\mu}{\lambda+\mu}\right)^{n} \left(\frac{n}{\mu}+\frac{1}{\lambda}\right)\right). \label{eq:app_MM1B_keep_old_ET}
    \end{split}
\end{equation}

We recall that the steady state probability $\pi_i$ of queue length being $i$ for $i\in\{0,1,...,B+1\}$ is given by 
\begin{equation}
    \pi_i = \frac{\rho^i}{\sum_{j=0}^{B+1} \rho^j} \text{ for } i = 0,1,\cdots,B+1, \label{eq:steady_prob}
\end{equation}
where $\rho = \frac{\lambda}{\mu}$ is the traffic intensity. Combining and rearranging $\mathbb{E}[S_k] = \frac{1}{\mu}$, (\ref{eq:app_MM1B_keep_old_EW}), (\ref{eq:app_MM1B_keep_old_ET}) and (\ref{eq:steady_prob}), we can obtain the long-term average peak age $\Bar{A}_{\text{Keep-Old}}(B)$ for an M/M/1/B+1 queueing system under the Keep-Old policy with a LCFS discipline as
\begin{equation}
    \begin{split}
        &\textstyle \Bar{A}_{\text{Keep-Old}}(B) = \frac{1}{\mu} + \frac{1}{\mathcal{C}_1} \Big[\frac{1}{\lambda}+\frac{1}{\mu-\lambda} \left(1+\frac{\lambda\mu}{(\lambda+\mu)^2}\right) \\
        &\hspace{2.8cm}\textstyle+\left(\frac{2}{\mu}-\frac{1}{\mu-\lambda}\left(1+\frac{\mu^2}{(\lambda+\mu)^2}\right)\right)\rho^B \Big], 
        \end{split}
\end{equation}
where $\mathcal{C}_1 = 1+\frac{\lambda\mu}{\mu^2-\lambda^2} - \frac{\lambda^2}{\mu^2-\lambda^2}\rho^B$ and  $\rho = \frac{\lambda}{\mu}$.

\section{Proof of Lemma~\ref{lemma:MM1B_keep_old_error}} \label{appendix:MM1B_keep_old_error}
Suppose that packet $i-1$ arrives under the event $\psi_n$, where $n=0,...B-1$. In this case, both packet $i-1$ and the next arriving packet $i$ will be served under the Keep-Old policy. Thus, the inter-arrival time $Z_i$ between these packets corresponds to the time elapsed between their arrivals. Since the arrival process follows a Poisson process with rate $\lambda$, the inter-arrival time $Z_i$ conditioned on $\psi_n$ is exponentially distributed with mean $\frac{1}{\lambda}$. Thus, the conditional second moment $\mathbb{E}[Z_i^2 | \psi_n]$ is given by
\begin{equation}
    \mathbb{E}[Z_i^2|\psi_n] = \frac{2}{\lambda^2} \text{ for } n = 0,...,B-1. \label{eq:app_MM1B_keep_old_Z2_n}
\end{equation}

Now, consider the case where packet $i-1$ arrives under the event $\psi_B$. At the arrival of packet $i-1$, the system has $B+1$ packets, including the in-service packet. Under the Keep-Old policy, any newly arriving packets during the remaining service time of the in-service packet will be dropped. After the departure of the in-service packet, the first arriving packet $i$ will be stored in the buffer. Consequently, the inter-arrival time $Z_i$ between packets $i-1$ and $i$ is composed of the two independent components: (1) the remaining service time of the in-service packet, which is exponentially distributed with rate $\mu$, and (2) the time interval between the departure of the in-service packet and the arrival of packet $i$, which is exponentially distributed with rate $\lambda$. Using the independence of these components and their distributions, the conditional second moment $\mathbb{E}[Z_i^2|\psi_B]$ is given by
\begin{equation}
    \mathbb{E}[Z_i^2 | \psi_B] = 2\left(\frac{1}{\mu^2}+\frac{1}{\lambda\mu}+\frac{1}{\lambda^2}\right). \label{eq:app_MM1B_keep_old_Z2_B}
\end{equation}

Combining (\ref{eq:app_MM1B_keep_old_Z2_n}) and (\ref{eq:app_MM1B_keep_old_Z2_B}), we have
\begin{equation}
    \mathbb{E}[Z_{\text{Keep-Old}}^2] = 2\pi_B \left(\frac{1}{\mu^2}+\frac{1}{\lambda\mu}+\frac{1}{\lambda^2}\right) + \frac{2}{\lambda^2} \sum_{n=0}^{B-1} \pi_n.
\end{equation}
Rearranging this with (\ref{eq:steady_prob}), we can obtain the long-term average reconstruction error $\overline{RE}_{\text{Keep-Old}}(B)$ for an M/M/1/B+1 queueing system under the Keep-Old policy with a LCFS discipline as
    \begin{equation}
        \overline{RE}_{\text{Keep-Old}}(B) = \frac{\lambda_{\text{eff}}\mathbb{E}[Z_{\text{Keep-Old}}^2]}{6},
    \end{equation}
    where
    \begin{equation}
    \lambda_{\text{eff}} = \frac{\lambda (1-\rho^{B+1})}{1-\rho^{B+2}},
    \end{equation}
    \begin{equation}
        \mathbb{E}[Z_{\text{Keep-Old}}^2] = \frac{2}{\mathcal{C}_2(\mu-\lambda)} \left(\frac{\mu}{\lambda^2}-\frac{\lambda}{\mu^2} \rho^B\right),
    \end{equation}
    and $\mathcal{C}_2 = \frac{\mu}{\mu-\lambda}-\frac{\lambda}{\mu-\lambda}\rho^B$.

\section{Proof of Lemma~\ref{lemma:MM1B_keep_fresh_age}}\label{appendix:MM1B_keep_fresh_age}

\begin{figure}
    \centering
    \includegraphics[width=0.9\linewidth]{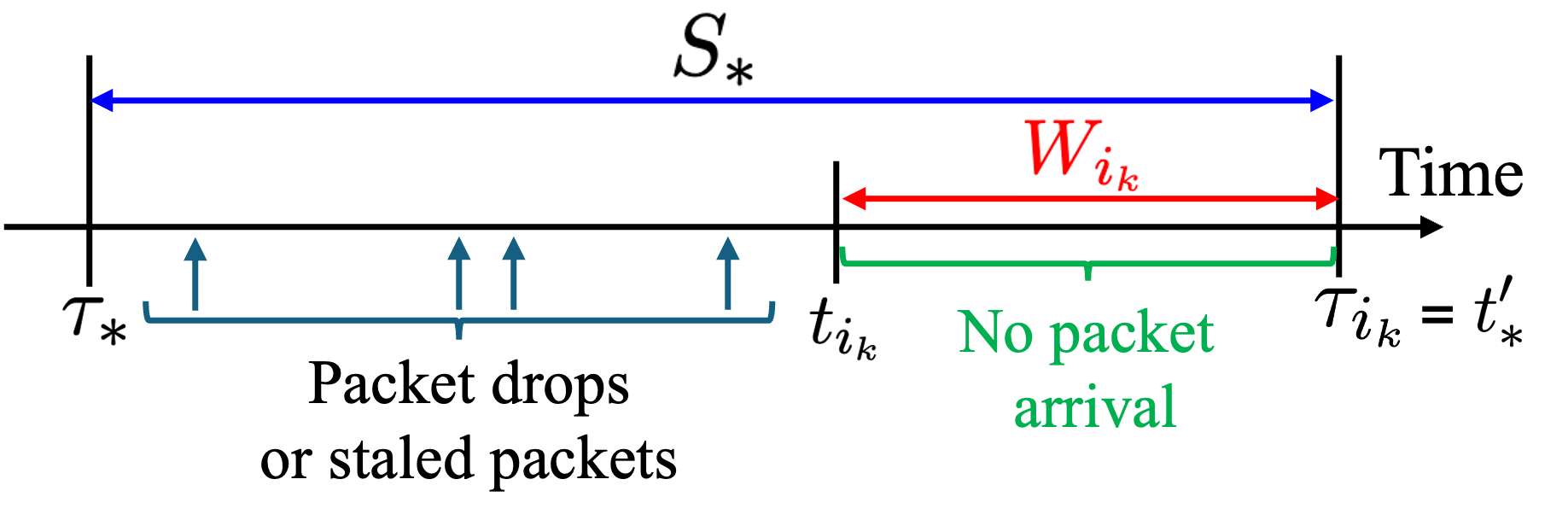}
    \caption{Sample path of packet arrivals and departures under the Keep-Fresh policy, where $\tau_*$ and $t'_*$ is the service start time and departure time of the in-service packet when packet $i_k$ arrives.}
    \label{fig:MM1B_keep_fresh_waiting_time}
\end{figure}

Suppose that packet $i_k$ arrives under the event $\psi_n$, where $n=1,...,B+1$ as shown in Fig.~\ref{fig:MM1B_keep_fresh_waiting_time}. Under the LCFS Keep-Fresh policy, if new arrivals occur during the remaining service time of the in-service packet, the earlier arriving packets will be staled.  Hence, packet $i_k$ remains fresh if no new arrival occurs during the remaining service time of the in-service packet, which happens with probability $\frac{\mu}{\lambda+\mu}$ from (\ref{eq:app_p_no_arrival}). Further, from (\ref{eq:app_pdf_conditional_w}), the conditional expected waiting time $\mathbb{E}[W|\psi_n,\text{fresh}]=\frac{1}{\lambda+\mu}$. Hence, the expected waiting time $\mathbb{E}[W_{i_k}]$ is given by
\begin{equation}
    \begin{split}
        \mathbb{E}[W_{i_k}] = \sum_{n=1}^{B+1} \frac{\mu\pi_n}{\lambda+\mu} \frac{1}{\lambda+\mu}.  \label{eq:app_MM1B_keep_fresh_EW}
    \end{split}
\end{equation}

For the inter-departure time $\mathbb{E}[Y_k]$. the conditional expectation $\mathbb{E}[Y_k|\psi_0]$ under the event $\psi_0$ matches the Keep-Old policy. This is because the system dynamics are identical when the buffer is empty, regardless of the policy. Thus, we have $\mathbb{E}[Y_k|\psi_0]=\frac{\lambda}{(\lambda+\mu)\mu}+\frac{1}{\lambda}$. If a packet arrives under the event $\psi_n$, where $n=1,...,B+1$, it is fresh with probability $\frac{\mu}{\lambda+\mu}$. Unlike the Keep-Old policy, packets arriving under the event $\psi_{B+1}$ replace the buffer's tail and may still be dropped if subsequent arrivals occur. The conditional expected inter-departure time $\mathbb{E}[Y|\psi_n,\text{fresh}]$ for $n=1,...,B$ remains identical to the Keep-Old policy due to the LCFS dynamics. Additionally, for the event $(\psi_{B+1},\text{fresh})$, the dynamics match $(\psi_{B},\text{fresh})$, so we have $\mathbb{E}[Y|\psi_{B+1},\text{fresh}]=\mathbb{E}[Y|\psi_{B},\text{fresh}]$. From (\ref{eq:app_MM1B_keep_old_conditional_ET}), the expected inter-departure time is given by (\ref{eq:MM1B_keep_fresh_ET}).

\begin{figure*}[!ht]
    \begin{equation}
    \begin{split}
        \mathbb{E}[Y_k] & = \pi_0 \left(\frac{\lambda}{(\lambda+\mu)\mu} + \frac{1}{\lambda}\right) + \frac{\mu}{\lambda+\mu}\sum_{n=1}^{B-1}\pi_n \left(\sum_{j=0}^{n-1} \left(\frac{\mu}{\lambda+\mu}\right)^{j} \frac{\lambda}{\lambda+\mu} \frac{j+1}{\mu} + \left(\frac{\mu}{\lambda+\mu}\right)^{n} \left(\frac{n}{\mu}+\frac{1}{\lambda}\right)\right) \\
        &\hspace{1cm}+ \frac{\mu}{\lambda+\mu}(\pi_B+\pi_{B+1})\left(\sum_{j=0}^{B-1} \left(\frac{\mu}{\lambda+\mu}\right)^{j} \frac{\lambda}{\lambda+\mu} \frac{j+1}{\mu} + \left(\frac{\mu}{\lambda+\mu}\right)^{B} \left(\frac{B}{\mu}+\frac{1}{\lambda}\right)\right).
        \label{eq:MM1B_keep_fresh_ET} 
    \end{split}
    \end{equation}
    \hrule
\end{figure*}

% \begin{strip}
%     \begin{equation}
%     \begin{split}
%         \mathbb{E}[Y_k] &= \pi_0 \left(\frac{\lambda}{(\lambda+\mu)\mu} + \frac{1}{\lambda}\right) \\
%         &+ \frac{\mu}{\lambda+\mu}\sum_{n=1}^{B-1}\pi_n \left(\sum_{j=0}^{n-1} \left(\frac{\mu}{\lambda+\mu}\right)^{j} \frac{\lambda}{\lambda+\mu} \frac{j+1}{\mu} + \left(\frac{\mu}{\lambda+\mu}\right)^{n} \left(\frac{n}{\mu}+\frac{1}{\lambda}\right)\right) \\
%         &+ \frac{\mu}{\lambda+\mu}(\pi_B+\pi_{B+1})\left(\sum_{j=0}^{B-1} \left(\frac{\mu}{\lambda+\mu}\right)^{j} \frac{\lambda}{\lambda+\mu} \frac{j+1}{\mu} + \left(\frac{\mu}{\lambda+\mu}\right)^{B} \left(\frac{B}{\mu}+\frac{1}{\lambda}\right)\right).
%         \label{eq:MM1B_keep_fresh_ET} 
%     \end{split}
%     \end{equation}
% \end{strip}

% \begin{figure*}[!ht]
%     \begin{equation}
%     \begin{split}
%         \mathbb{E}[Y_k] &= \pi_0 \left(\frac{\lambda}{(\lambda+\mu)\mu} + \frac{1}{\lambda}\right) + \frac{\mu}{\lambda+\mu}\sum_{n=1}^{B-1}\pi_n \left(\sum_{j=0}^{n-1} \left(\frac{\mu}{\lambda+\mu}\right)^{j} \frac{\lambda}{\lambda+\mu} \frac{j+1}{\mu} + \left(\frac{\mu}{\lambda+\mu}\right)^{n} \left(\frac{n}{\mu}+\frac{1}{\lambda}\right)\right) \\
%         &\hspace{1cm}+ \frac{\mu}{\lambda+\mu}(\pi_B+\pi_{B+1})\left(\sum_{j=0}^{B-1} \left(\frac{\mu}{\lambda+\mu}\right)^{j} \frac{\lambda}{\lambda+\mu} \frac{j+1}{\mu} + \left(\frac{\mu}{\lambda+\mu}\right)^{B} \left(\frac{B}{\mu}+\frac{1}{\lambda}\right)\right).
%         \label{eq:MM1B_keep_fresh_ET} 
%     \end{split}
%     \end{equation}
%     \hrule
% \end{figure*}

Combining and rearranging $\mathbb{E}[S_{i_k}]=\frac{1}{\mu}$, (\ref{eq:app_MM1B_keep_fresh_EW}), (\ref{eq:MM1B_keep_fresh_ET}) and (\ref{eq:steady_prob}), we can obtain the long-term average peak age $\Bar{A}_{\text{Keep-Fresh}}(B)$ for an M/M/1/B+1 queueing system  under the Keep-Fresh policy with a LCFS discipline as
\begin{align}
    \Bar{A}_{\text{Keep-Fresh}}(B) &= \frac{1}{\mu} + \frac{1}{\mathcal{C}_1} \bigg[\frac{1}{\lambda}+\frac{1}{\mu-\lambda} \left(1+\frac{\lambda\mu}{(\lambda+\mu)^2}\right) \nonumber\\
    &+ \left(\frac{1}{\mu}-\frac{1}{\mu-\lambda}\left(1+\frac{\lambda^2}{(\lambda+\mu)^2}\right)\right)\rho^B\bigg],
\end{align}
where $\mathcal{C}_1 = 1+\frac{\lambda\mu}{\mu^2-\lambda^2} - \frac{\lambda^2}{\mu^2-\lambda^2}\rho^B$ and  $\rho = \frac{\lambda}{\mu}$.

\section{Proof of Lemma~\ref{lemma:MM1B_keep_fresh_error}}
\label{appendix:MM1B_keep_fresh_error}

\begin{figure}[t]
    \centering
    \begin{subfigure}[b]{0.47\textwidth}
        \centering
        \includegraphics[width=0.7\linewidth]{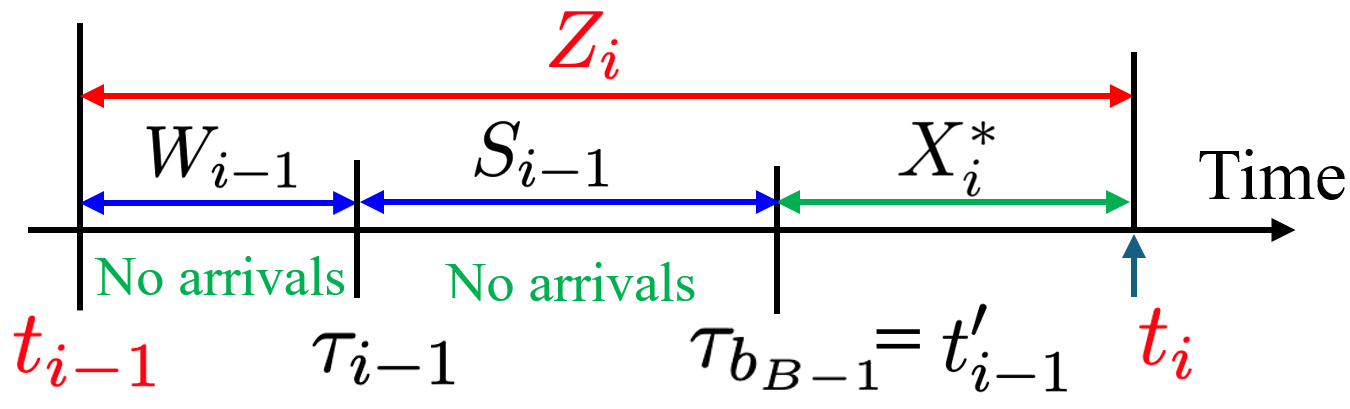}
        \caption{No arrival occurs during the service time $S_{i_{k-1}}$.}     
        \label{subfig:MM1B_keep_fresh_re_3_appendix}
    \end{subfigure}
    \hfill
    \begin{subfigure}[b]{0.48\textwidth}
        \centering     
        \includegraphics[width=0.7\linewidth]{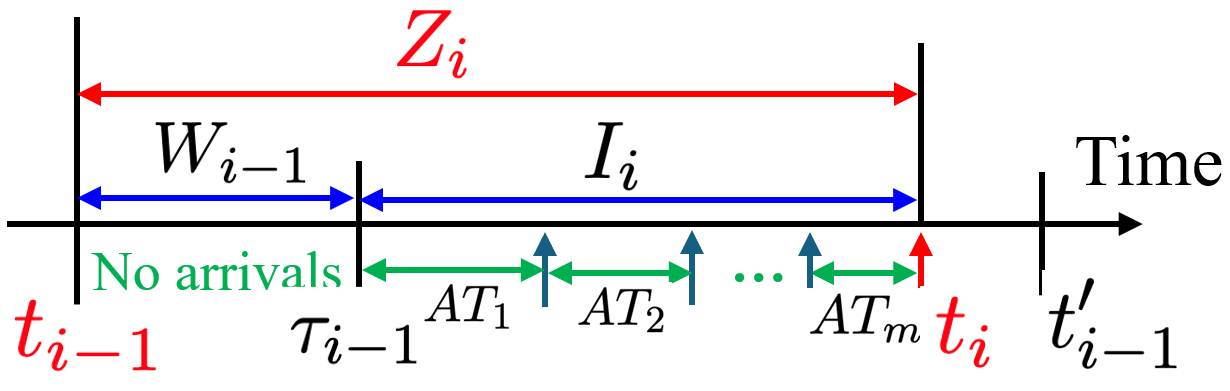}
        \caption{New arrivals occur during the service time $S_{i_{k-1}}$.}
        \label{subfig:MM1B_keep_fresh_re_4_appendix}
    \end{subfigure}
    \caption{Sample path of packet arrivals and departures under the Keep-Fresh policy, where $t'_*$ is the departure time of the in-service packet.}
    \label{fig:MM1B_keep_fresh_re2_appendix}
\end{figure}

Packets arriving under the event $\psi_n$, where $n = 0, ..., B-1$, are guaranteed to be served because the buffer is not full upon their arrival. However, packets arriving under $\psi_B$ or $\psi_{B+1}$ may either be served or dropped, depending on whether fresher packets arrive before the current in-service packet completes its service.

If packet $i-1$ arrives under the event $\psi_n$, where $n=0,...B-2$, both packet $i-1$ and the next arriving packet $i$ will be served. In this scenario, the inter-arrival time $Z_i$ between packets $i-1$ and $i$ is unaffected by the policy, as no additional arrivals are dropped. Thus, following the same line of reasoning as in the derivation for the Keep-Old policy, the conditional second moment $\mathbb{E}[Z_i^2 | \psi_n]$ is given by
\begin{equation}
    \mathbb{E}[Z_i^2|\psi_n] = \frac{2}{\lambda^2} \text{ for } n = 0,...,B-2. \label{eq:MM1B_keep_fresh_Z2_n}
\end{equation}

Suppose that packet $i-1$ arrives under the event $\psi_{B-1}$. At this point, the system contains $B$ packets, including the in-service packet, leaving one remaining slot in the buffer. In this case, the inter-arrival time $Z_i$ between packets $i-1$ and $i$ depends on whether new arrivals occur during the remaining service time of the in-service packet ($\bar{\phi}_R$) or not ($\phi_R$). The dynamics of $Z_i$ under these conditions are equivalent to the dynamics of $I_i$ in the $M/M/1/2$ queue analyzed in Proof of Lemma III.2, due to the memoryless property of service times. We recall the following result:
\begin{align}
    \mathbb{E}[I_i]
    %&= \mathbb{P}(\phi)\mathbb{E}[I_i|\phi]+\sum_{m=1}^\infty \mathbb{P}(\bar{\phi}_{s,m})\mathbb{E}[I_i|\bar{\phi}_{s,m}] \nonumber\\
    %&= \frac{\mu}{\lambda+\mu}\frac{2\lambda+\mu}{\lambda(\lambda+\mu)} + \sum_{m=1}^\infty \left(\frac{\lambda}{\lambda+\mu}\right)^{m} \frac{\mu}{\lambda+\mu} \frac{m}{\lambda+\mu} \nonumber \\
    &=  \frac{\mu(2\lambda+\mu)}{\lambda(\lambda+\mu)^2} + \frac{\lambda}{\mu(\lambda+\mu)}. \label{eq:app_EI_total}
\end{align}
\begin{align} 
\mathbb{E}[I^2_i]
%&= \mathbb{P}(\phi)\mathbb{E}[I^2_i|\phi]+\sum_{m=1}^\infty \mathbb{P}(\bar{\phi}_{s,m})\mathbb{E}[I^2_i|\bar{\phi}_{s,m}] \nonumber\\
    %&= \frac{\mu}{\lambda+\mu} \frac{2(3\lambda^2+3\lambda\mu+\mu^2)}{\lambda^2(\lambda+\mu)^2} \nonumber \\
    %&\hspace{1.8cm}+ \sum_{m=1}^\infty \left(\frac{\lambda}{\lambda+\mu}\right)^{m} \frac{\mu}{\lambda+\mu} \frac{m(m+1)}{(\lambda+\mu)^2} \nonumber \\
    &= \frac{2\mu(3\lambda^2+3\lambda\mu+\mu^2)}{\lambda^2(\lambda+\mu)^3}+\frac{2\lambda}{\mu^2(\lambda+\mu)}. \label{eq:app_EI2_total}
\end{align}
Therefore, from (\ref{eq:app_EI2_total}), the conditional second moment of $Z_i$ is given by
\begin{equation}
    \begin{split}
        \mathbb{E}[Z_i^2|\psi_{B-1}] = \frac{2\mu(3\lambda^2+3\lambda\mu+\mu^2)}{\lambda^2(\lambda+\mu)^3} + \frac{2\lambda}{\mu^2 (\lambda+\mu)}. \label{eq:app_MM1B_keep_fresh_Z2_B1}
    \end{split}
\end{equation}

Lastly, consider a packet arriving under the event $\psi_B$ or $\psi_{B+1}$. At the time of its arrival, the system is full, so the packet is not guaranteed to be served. From (\ref{eq:app_p_no_arrival}), the probability of its transmission is given by $\mathbb{P}(\text{tx}|\psi_B \text{ or } \psi_{B+1}) = \mathbb{P}(\phi_r) = \frac{\mu}{\lambda+\mu}$. 
Under the event $(\psi_B \text{ or } \psi_{B+1},\text{tx})$, suppose the event $\phi_s$ where no arrival occurs during the service time $S_{i-1}$ with probability $\frac{\mu}{\lambda+\mu}$, as shown in Fig.~\ref{fig:MM1B_keep_fresh_re2_appendix}(\subref{subfig:MM1B_keep_fresh_re_3_appendix}). In this case, after the departure of packet $i-1$, the system contains $B-1$ packets, meaning the next arriving packet $i$ encounters the event $\psi_{B-1}$, which guarantees its service. Thus, the inter-arrival time between packets $i-1$ and $i$ is composed of the waiting time $W_{i-1}$, the service time $S_{i-1}$ and the time interval $X^*_i$. From (\ref{eq:pdf_W}) and (\ref{eq:app_pdf_S}), $W_{i-1}$ and $S_{i-1}|\phi_s$ follow exponential distributions with rate $\lambda+\mu$. Further, $X^*_i$ follows an exponential distribution with rate $\lambda$. Since those intervals are independent each other, the conditional second moment of inter-arrival time $\mathbb{E}[Z^2_i | \psi_B \text{ or } \psi_{B+1}, \text{tx}, \phi_s]$ is given by
\begin{equation}
    \textstyle\mathbb{E}[Z^2_i | \psi_B \text{ or } \psi_{B+1}, \text{tx}, \phi_s] = \frac{6}{(\lambda+\mu)^2} + \frac{2}{\lambda^2} + \frac{4}{\lambda(\lambda+\mu)}. \label{eq:app_EZ2_B_0}
\end{equation}

Now, suppose the event $\bar{\phi}_s$ that new arrivals occur during the service time $S_{i-1}$ as shown in Fig.~\ref{fig:MM1B_keep_fresh_re2_appendix}(\subref{subfig:MM1B_keep_fresh_re_4_appendix}), which happens with probability $\frac{\lambda}{\lambda+\mu}$. In this case, the next arriving packet encounters the event $\psi_B$ and after-arriving packets encounter the event $\psi_{B+1}$. Thus, the latest arriving packet $i$ will be served after the departure of packet $i-1$. Hence, the inter-arrival time between packets $i-1$ and $i$ is composed of the waiting time time $W_{i-1}$ and the time interval $I_i$ between the departure of in-service packet and the arrival of packet $i$. From (\ref{eq:pdf_W}), (\ref{eq:app_EI_total}) and (\ref{eq:app_EI2_total}), the conditional expected inter-arrival time $\mathbb{E}[Z^2_i | \psi_B \text{ or } \psi_{B+1}, \text{tx}, \bar{\phi}_s]$ is given by 
\begin{align}
    &\mathbb{E}[Z^2_i | \psi_B \text{ or } \psi_{B+1},\text{tx}, \bar{\phi}_s] \nonumber\\
    &\hspace{0.4cm}= \mathbb{E}[W_{i-1}^2|\cdot] + 2\mathbb{E}[W_{i-1}|\cdot] \mathbb{E}[I_i|\cdot] + \mathbb{E}[I_i^2|\cdot] \nonumber\\
    &\hspace{0.4cm}= \frac{2}{(\lambda+\mu)^2} + \frac{2}{\lambda+\mu} \sum_{m=1}^{\infty} \left(\frac{\lambda}{\lambda+\mu}\right)^{m-1} \frac{\mu}{\lambda+\mu} \frac{m}{\lambda+\mu} \nonumber\\
    &\hspace{0.8cm}+ \sum_{m=1}^{\infty} \left(\frac{\lambda}{\lambda+\mu}\right)^{m-1} \frac{\mu}{\lambda+\mu} \frac{m(m+1)}{(\lambda+\mu)^2}\nonumber\\
    &\hspace{0.4cm}= \frac{2}{(\lambda+\mu)^2} + \frac{2}{\lambda+\mu} \frac{1}{\mu} + \frac{2}{\mu^2}, \label{eq:app_EZ2_B_1}
\end{align} 
where we omit the condition ``$\psi_B \text{ or } \psi_{B+1}, \text{tx}, \bar{\phi}_s$'' in the first line due to the limited space. 

Combining (\ref{eq:app_EZ2_B_0}) and (\ref{eq:app_EZ2_B_0}), we have
\begin{align}
    &\mathbb{E}[Z^2_i | \psi_B \text{ or } \psi_{B+1},\text{tx}] \textstyle= \frac{\mu}{\lambda+\mu} \left(\frac{6}{(\lambda+\mu)^2} + \frac{2}{\lambda^2}
    + \frac{4}{\lambda(\lambda+\mu)}\right) \nonumber \\
    &\hspace{3.7cm}\textstyle+ \frac{\lambda}{\lambda+\mu} \left(\frac{2}{(\lambda+\mu)^2} + \frac{2}{\lambda+\mu} \frac{1}{\mu} + \frac{2}{\mu^2}\right) \nonumber \\
    &\hspace{0.4cm}= \frac{2\mu^3(6\lambda^2+4\lambda\mu+\mu^2)+2\lambda^3(\lambda^2+3\lambda\mu+3\mu^2)}{\lambda^2 \mu^2 (\lambda+\mu)^3}. \label{eq:app_MM1B_keep_fresh_Z2_B2}
\end{align}
Finally, combining (\ref{eq:MM1B_keep_fresh_Z2_n}), (\ref{eq:app_MM1B_keep_fresh_Z2_B1}) and (\ref{eq:app_MM1B_keep_fresh_Z2_B2}), we have
\begin{align}
    &\mathbb{E}[Z_{\text{Keep-Fresh}}^2]\nonumber\\
    &= \frac{2}{\lambda^2} \sum_{n=0}^{B-2} \pi_n + \pi_{B-1} \left(\frac{2\mu(3\lambda^2+3\lambda\mu+\mu^2)}{\lambda^2(\lambda+\mu)^3} + \frac{2\lambda}{\mu^2 (\lambda+\mu)}\right) \nonumber\\
    &\hspace{0.4cm}+ (\pi_B + \pi_{B+1})\frac{\mu}{\lambda+\mu} \mathbb{E}[Z^2_k | \psi_B \text{ or } \psi_{B+1},\text{tx}].
\end{align}

Rearranging this, we can obtain the long-term average reconstruction error $\overline{RE}_{\text{Keep-Fresh}}(B)$ for an M/M/1/B+1 queueing system under the Keep-Fresh policy with a LCFS discipline as
    \begin{equation}
        \overline{RE}_{\text{Keep-Fresh}}(B) = \frac{\lambda_{\text{eff}}\mathbb{E}[Z_{\text{K-F}}^2]}{6},
    \end{equation}
    where
    \begin{equation}
    \lambda_{\text{eff}} = \frac{\lambda (1-\rho^{B+1})}{1-\rho^{B+2}}, \text{ and }\nonumber
    \end{equation}
    \begin{equation}
        \mathbb{E}[Z_{\text{K-F}}^2] = \frac{1}{\mathcal{C}_2}\left( \frac{2\mu}{(\mu-\lambda)\lambda^2}+ \left(\mathcal{Z}_1 + \mathcal{Z}_2 - \frac{2\mu^2}{(\mu-\lambda)\lambda^3}\right)\rho^B\right), \nonumber
    \end{equation}
    where
    \begin{align}  
        \mathcal{Z}_1 &= \frac{2\mu^3(6\lambda^2+4\lambda\mu+\mu^2)+2\lambda^3(\lambda^2+3\lambda\mu+3\mu^2)}{\lambda^2 \mu^2 (\lambda+\mu)^3},\\
        \mathcal{Z}_2 &= \frac{2\mu^2(3\lambda^2+3\lambda\mu+\mu^2)}{\lambda^3(\lambda+\mu)^3} + \frac{2}{\mu(\lambda+\mu)},
    \end{align}
    and $\mathcal{C}_2 = \frac{\mu}{\mu-\lambda}-\frac{\lambda}{\mu-\lambda}\rho^B$.

\end{document}